\documentclass[prd,aps,nofootinbib,showpacs,preprintnumbers,10pt,twocolumn]{revtex4}
\usepackage{amsmath,amssymb,graphicx,epsfig,float}

\begin{document}

\title{$B\to DK^*_{0,2}$ Decays:\\
PQCD analysis  to determine  CP violation phase angle $\gamma$}
\author{C.S. Kim$^1$\footnote{Email: cskim@yonsei.ac.kr}, Run-Hui Li$^1$\footnote{Email: lirh@yonsei.ac.kr}, Wei Wang$^2$\footnote{Email: weiwang@hiskp.uni-bonn.de}}
\affiliation{$^1$ Department of Physics $\&$ IPAP, Yonsei
University, Seoul 120-479, Korea \\
$^2$Helmholtz-Institut f\"ur Strahlen- und Kernphysik and Bethe Center for Theoretical Physics,
Universit\"at Bonn, D-53115 Bonn, Germany}

\begin{abstract}
\noindent $B^\pm \to (D^0, \bar D^0, D_{CP})K^{*\pm}_{0,2}$  decays  are helpful in determining the CP violation angle $\gamma$, and
we analyze  these decay processes
within the perturbative QCD approach based on $k_T$ factorization. We found that the branching ratio of $B^-\to D^0 K^{*-}_0$ can reach the
order of  $10^{-4}$, due to the enhancement of nonfactorizable
contributions in color-suppressed $D^0$-emission, while  the
branching ratio of $B^-\to \bar D^0K^{*-}_0$ is  of the order
$10^{-5}$. The ratio of  decay
amplitudes is about 3 times larger than the one in the channel
$B^\pm\to DK^\pm$.  Large branching
ratios provide a good opportunity to observe $ B^\pm \to D K^{*\pm}_2$ on
the ongoing and forthcoming experimental facilities and consequently
these channels may be of valuable avail in reducing the errors in the CP violation
phase angle $\gamma$. We also explore the possible time-dependent CP
asymmetries of $B_s$ decay into a scalar meson to determine the phase angle $\gamma$.
\end{abstract}

\maketitle

%%%%%%%%%%%%%%%%%%%%%%%%%%%%%%%%%%%%%%%%%%%%%%%%%%%%%%%%%%%%%%%%%%%%
\section{Introduction}
%%%%%%%%%%%%%%%%%%%%%%%%%%%%%%%%%%%%%%%%%%%%%%%%%%%%%%%%%%%%%%%%%%%%
%\section{Introduction}

The authentication of the unitarity of CKM matrix  allows us to
explore the  standard model (SM) description of the CP violation and
reveal new physics beyond the SM. Among the angles
$(\alpha,\beta,\gamma)$ of the  so-called $(bd)$ unitarity triangle derived from the $V_{ud}V_{ub}^*+V_{cd}V_{cb}^*+V_{td}V_{tb}^*=0$,
satisfying  the constraint $\alpha+\beta+\gamma=180^\circ$, the
angle $\gamma$ are least constrained, with a precision of roughly
$10^\circ$. This is one of  the main sources of the current
uncertainties in the apex of the unitary
triangle~\cite{Asner:2010qj,Lees:2013zd}.

One of the most efficient ways proposed in the literature to measure
$\gamma$ makes use of the two triangles formed by the six channels of
$B^\pm\to (D^0,\bar D^0, D_{CP})
K^\pm$~\cite{Gronau:1991dp,Gronau:1990ra,Dunietz:1991yd}. The shape
of the two triangles is controlled  by two quantities
\begin{eqnarray}
 r_{B}^{K_J}\equiv\left|{A(B^-\to \bar D^0 {K_J^-})}/{A(B^-\to D^0 K^{-}_J)}\right|,\nonumber\\
 \delta_{B}^{K_J} \equiv arg\left[{e^{i\gamma} A(B^-\to \bar D^0 K^{-}_J)}/{A(B^-\to D^0 K^{-}_J)}\right],\nonumber
\end{eqnarray}
where $K_J$ can be $ K$ or  $K^*_{0,2}$.  One of the most intriguing properties in this method  is that it  is independent
of hadronic uncertainties, and moreover the CP violation from the $D$ meson decays can also be incorporated~\cite{Wang:2012ie}.  Due to the fact that the $B^-\to \bar D^0 {K^-}$ is
both Cabibbo-suppressed and color suppressed, the ratio  $r_B^K \sim
|V_{ub}V_{cs}^*/(V_{cb}V_{us}^*) a_2/a_1|\sim 0.1$ is small and in
particular the world averages for these parameters~\cite{CKMfitter}
\begin{eqnarray}
 r_B^K= 0.107\pm 0.010,\;\;
 \delta_B^K=( 112^{+12}_{-13})^\circ \nonumber
\end{eqnarray}
indicate that the two triangles formed by decay amplitudes are squashed. As a consequence,  the
measurement of $\gamma$ requests a precise knowledge on the $B^-\to
\bar D^0 {K^-}$.

In Ref.~\cite{Wang:2011zw}, we proposed a new method to determine
the CP violation angle $\gamma$ that uses the $B^\pm \to
D K^{*\pm}_{0,2}$ decays (see also Ref.~\cite{Diehl}).  Unlike the $B\to DK^\pm$,   the
color-allowed amplitudes in $B^\pm \to D K^{*\pm}_{0,2}$ have
vanishing/small decay constants and are comparable with the
color-suppressed ones. Large interference between the two amplitudes is induced in the $B\to D_{CP}K$ and  the
sensitivity to $\gamma$ is greatly  improved.  Branching ratios of
these channels are estimated to  lie in the range from  $10^{-6}$ to
$10^{-5}$, using a method of factorization in conjunction with
experimental data~\cite{Wang:2011zw}.  The motif of this work is to adopt  the
QCD-based factorization method, more explicitly the perturbative QCD (PQCD)
approach~\cite{Li:1994iu,Keum:2000wi,Lu:2000em} (see Ref.~\cite{Li:2012nk} and Ref.~\cite{Wang:2012ab} for the recent developments and applications of the PQCD approach), to calculate the
branching ratios, strong phases and CP asymmetries. The perturbative QCD approach is formulated
on the basis of  $k_T$ factorization, and   has been
applied  to $B$ meson decays into charmed meson in a number of
references and a global agreement of the results with the available
data is
found~\cite{Kurimoto:2002sb,Keum:2003js,Lu:2002iv,Li:2003wg,Lu:2003xc,Li:2008ts,Zou:2009zza,Li:2009xf}.  One of the most
successful predictions  is $r_K=
0.092^{+0.012+0.003}_{-0.003-0.003}$~\cite{Zou:2009zza}, in good
agreement with the data~\cite{CKMfitter}.
To the end of this work, we show that the resulting branching ratios
are enhanced by one order of magnitude than our previous estimates,
due to the inclusion  of the large nonfactorizable contribution in
$D^0$ emission diagram.  Such large branching ratios provide a
better opportunity  for the measurement  of  these channels on the experimental facilities  and
constraining  the $\gamma$ angle.

The rest of this work is organized as follows: In
Sec.~\ref{sec:K*0}, we  will calculate the  $B \to D^0(\bar D^0)
K^{*}_{0,2}(1430)$ decay amplitudes and give the factorization
formulas, while  Sec.~\ref{sec:f0} contains the numerical
analysis and discussions. The last section is our summary. We also relegate some of the calculation details to the Appendix.

\section{perturbative QCD Calculation  }\label{sec:K*0}

In the PQCD approach, the inclusion of the intrinsic transverse momentum of valence quarks  smears the endpoint singularities appearing in the calculations under the collinear factorization context.
In the $m_b\to \infty$ limit, the decay amplitude is generically expressed
as a convolution of wave functions and hard
scattering kernel with both longitudinal momenta and
transverse space coordinates
\begin{eqnarray}
{\cal M}=\int^1_0dx_1dx_2 dx_3\int
{d^2{\vec{b}}_{1}}{d^2{\vec{b}}_{2}}{d^2{\vec{b}}_{3}}{\phi}_B(x_1,{\vec{b}}_{1},t) \nonumber\\
\times T_H(x_1,x_2,{\vec{b}}_{1},{\vec{b}}_{2},t){\phi}_2(x_2,{\vec{b}}_{2},t){\phi}_3(x_3,{\vec{b}}_{3},t),
\end{eqnarray}
where the $B$ in the indices represents a $B$ meson and $2,3$ represent the two mesons in the final state.
In the computation of higher order QCD corrections,
the overlap  of soft
and collinear momentum results in double logarithm divergences. Resummation of them leads to the Sudakov factor which has the tendency to diminish the endpoint contributions and supports the hard-scattering picture used in this framework.  For a review of this approach, see Ref.~\cite{Li:2003yj}.

The wave functions, the most important entry in the perturbative QCD approach, are nonperturbative in nature and can only be acquired by some nonperturbative methods or with the aid from some simple but effective models.
For the $B$ meson which is a heavy-light system, we adopt the   light cone matrix
 \begin{equation}
 \Phi_{B}=\frac{i}{\sqrt{2N_c}}( p\!\!\!\slash_B +m_B)\gamma_5\phi_{B} (x_1,b_1), \label{Bwave:3variable}
 \end{equation}
 in which we have neglected the numerically-suppressed distribution amplitude~\cite{Lu:2002ny}.
Here $x_1$ is the momentum fraction of the light spectator quark and
$N_c=3$ is the color factor.
As for the wave functions for the $D$ meson, we use the form  derived in Ref.~\cite{Kurimoto:2002sb}
\begin{eqnarray}
 \Phi_{D}= \frac{i}{\sqrt {2N_c}} \gamma_5 (p\!\!\!\slash_D+m_D)\phi_{D}(x_2,b_2).
\end{eqnarray}

The light-cone distribution amplitudes (LCDAs) for $K^*_0$ are governed  by the conformal spin symmetry of QCD and  have the following definitions~\cite{Cheng:2005nb}
 \begin{eqnarray}
 &&\langle K^*_0(p_{K^*_0})|q(0)_j\bar q(z)_l |0\rangle
 =\frac{-1}{\sqrt{2N_c}}\int^1_0dxe^{ixp_{K^*_0}\cdot
 z} \nonumber\\
&&\{ p\!\!\!\slash_{K^*_0}\phi_{K^*_0}(x)
 +m_{K^*_0}\phi^s_{{K^*_0}}(x)+m_{K^*_0}(\bar n\!\!\!\slash n\!\!\!\slash-1)\phi^T_{K^*_0}(x)\}_{jl},\nonumber
 \end{eqnarray}
in which $\bar n$ is chosen as the flight direction of the $K^*_0$ in the $B$ meson rest frame and $n$ is  opposite to $\bar n$.
These LCDAs,  the twist-2 $\phi_{K^*_0}$  and the  twist-3 $\phi_{K^*_0}^{s,T}$,   can be expanded in terms of Gegenbauer polynomials
 \begin{eqnarray}
 \phi_{K^*_0}(x)
 &=&\frac{\bar
 f_{K^*_0}}{2\sqrt{2N_c}}6x(1-x)\sum_{m=0}^{\infty}B_mC_m^{{3/2}}(2x-1),\nonumber\\
 \phi_{K^*_0}^s(x)&=&\frac{\bar f_{K^*_0}}{2\sqrt{2N_c}},\;\;
 \phi_{K^*_0}^T(x)= \frac{\bar f_{K^*_0}}{2\sqrt{2N_c}}(1-2x),
 \end{eqnarray}
with $B_0= ({m_s-m_u})/{m_{K^*_0}}$. The decay constant $\bar f_{K^*_0}$ is defined by a scalar current
\begin{eqnarray}
 \langle K^{*-}_0(1430)|\bar s u|0\rangle =\bar  f_{K^*_0},\nonumber
\end{eqnarray}
and is related to the vector decay constant by $f_{K^*_0}= B_0 \bar f_{K^*_0}$.
We will leave out  the higher Gegenbauer moments in twist-3 LCDAs~\cite{Lu:2006fr,Han:2013zg}
since their contributions are found to be typically small~\cite{Li:2008tk}.

Similarly the LCDAs of a longitudinally polarized $K^*_2$ state are defined as~\cite{Cheng:2010hn}
\begin{eqnarray}
&&\langle K^*_2(p_{K^*_2},\epsilon)|\bar q_{2\beta}(z) q_{1\alpha} (0)|0\rangle
=\frac{1}{\sqrt{2N_c}}\int_0^1 dx e^{ixp_{K^*_2}\cdot z}  \nonumber\\
&&\times \Big\{m_{K^*_2}\not\!
\epsilon^*_{\bullet L} \phi_{K^*_2}(x) +\not\! \epsilon^*_{\bullet
L}p\!\!\!\slash_{K^*_2} \phi_{{K^*_2}}^{t}(x) \nonumber\\
&& +m_{K^*_2}^2\frac{\epsilon_{\bullet L} \cdot
v}{p_{K^*_0}\cdot v} \phi_{K^*_2}^s(x)\Big\}_{\alpha\beta}, \label{eq:LCDAK*2}
\end{eqnarray}
with $n^2=v^2=0$ being light-like unit vectors.
The new vector $\epsilon_{\bullet L}$ in
Eq.~\eqref{eq:LCDAK*2} is related to the polarization tensor by
$\epsilon_{\bullet L\mu}\equiv\frac{\epsilon_{\mu\nu} v^\nu}{p_{K^*_2}\cdot
v}m_{K^*_2}$ and
can be simplified in terms of a polarization vector
\begin{eqnarray}
\epsilon_{\bullet L\mu}\equiv\frac{\epsilon_{\mu\nu} v^\nu}{p_{K^*_2}\cdot
v}m_{K^*_2}\simeq \frac{\sqrt {2} }{\sqrt 3} \epsilon_{L\mu}.
\end{eqnarray}
The above LCDAs have the asymptotic forms~\cite{Cheng:2010hn}
\begin{eqnarray}
&&\phi_{K^*_2}(x)=\frac{f_{K^*_2}}{2\sqrt{2N_c}} 30x(1-x)(2x-1),\nonumber\\
&&
\phi_{K^*_2}^t(x)=\frac{f_{K^*_2}^T}{2\sqrt{2N_c}}\frac{15}{2}(2x-1)(1-6x+6x^2),\nonumber\\
&&\phi_{K^*_2}^s(x)=\frac{f_{K^*_2}^T}{4\sqrt{2N_c}}
\frac{d}{dx}\left[15x(1-x)(2x-1)\right].
\end{eqnarray}

%%%%%%%%%%%%%%%%%%%%%%%%%%%%%%%%%%%%%%%%%%%%%%%%%%%%
\begin{figure}\begin{center}
\includegraphics[scale=0.5]{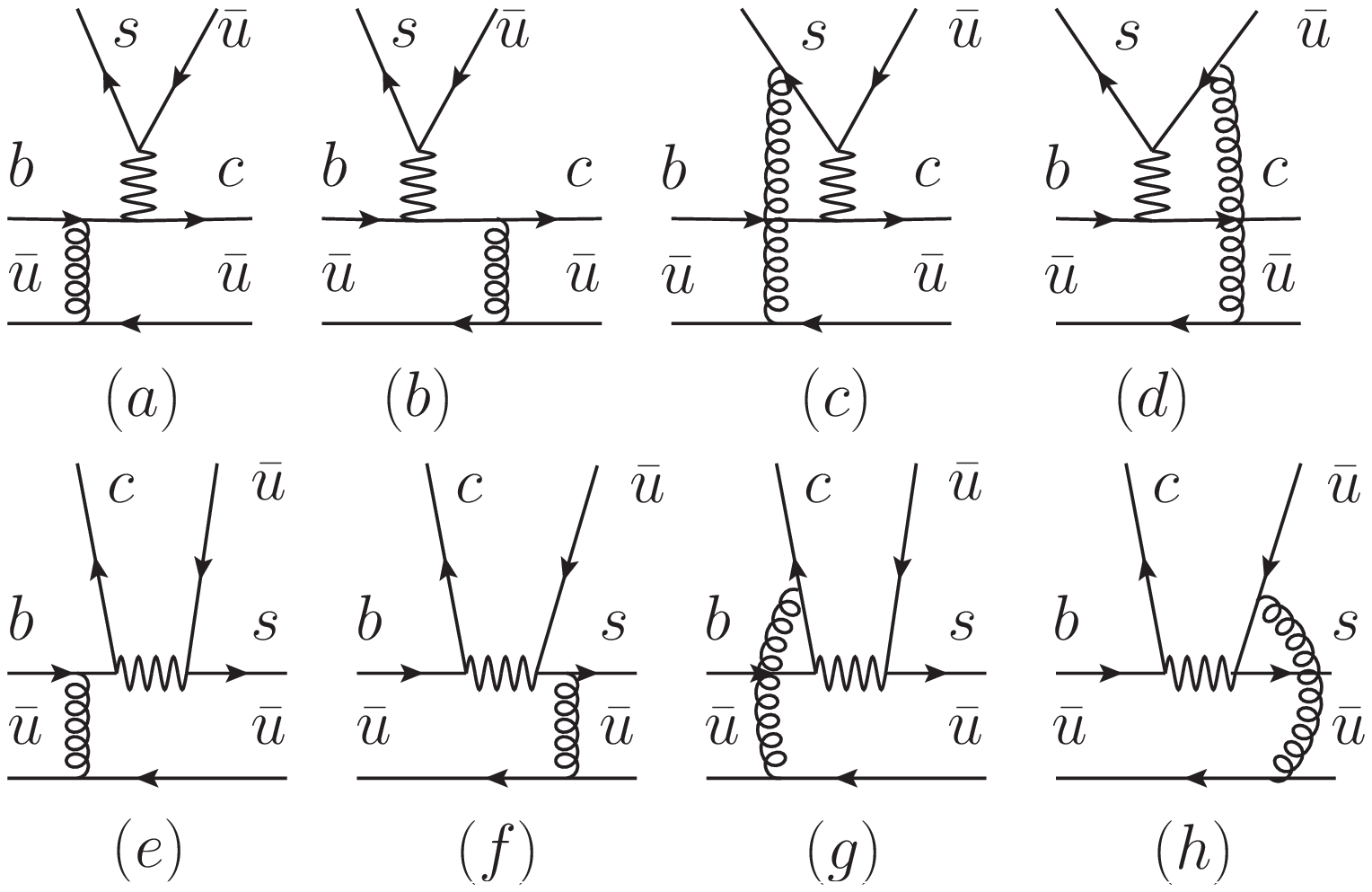}
\includegraphics[scale=0.35]{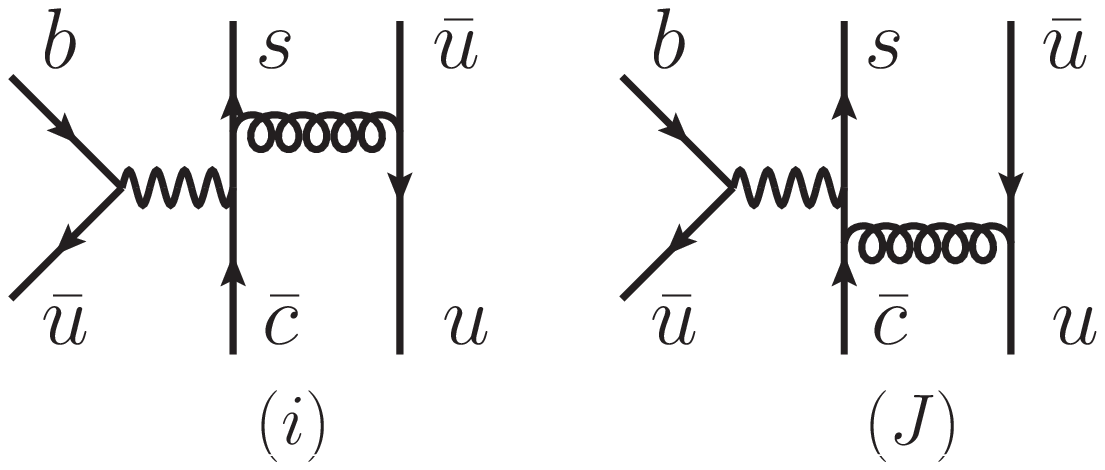}
\includegraphics[scale=0.35]{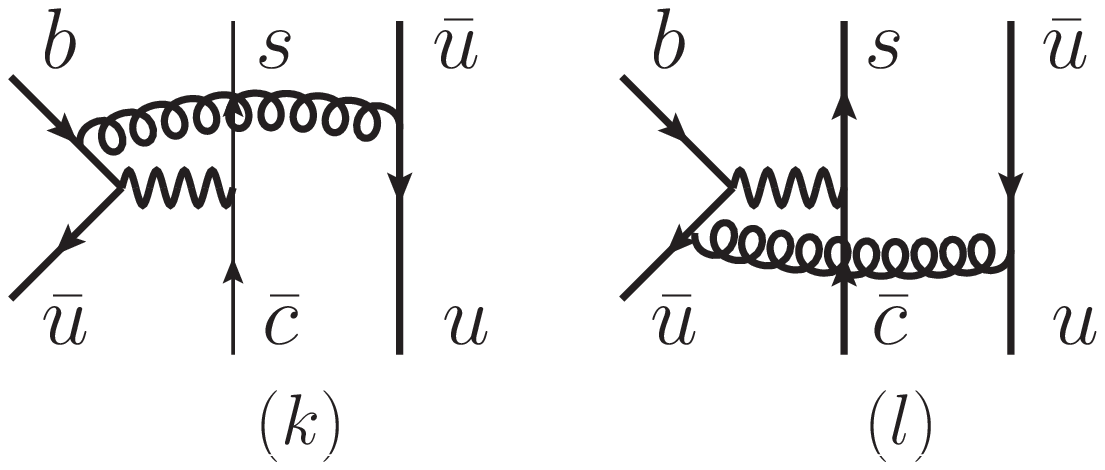}
\caption{Feynman diagrams for the color-allowed   contributions in
the process $B^-\to D^0 K^{*-}_{0(2)}(1430)$: (a,b,c,d), for the
color-suppressed   contributions in the process $B^-\to D^0
K^{*-}_{0(2)}(1430): (e,f,g,h)$, and for the annihilation contributions in process $B^-\to\bar D^0
K^{*-}_{0(2)}(1430): (i,j,k,l)$. In the middle four diagrams, the
replacement of the $c\bar u$ by $u\bar c$ results in the
corresponding diagrams for the process $B^-\to \bar D^0
K^{*-}_{0(2)}(1430)$. } \label{fig:colorallowed}
\end{center}
\end{figure}
%%%%%%%%%%%%%%%%%%%%%%%%%%%%%%%%%%%%%%%%%%%%%%%%%%%%%

There are three types of diagrams contributing  to the decay amplitudes which are depicted in Fig.~\ref{fig:colorallowed}:  the color-allowed contributions in the process $B^-\to D^0 K^{*-}_{0(2)}(1430)(a,b,c,d)$, the color-suppressed one in the process $B^-\to D^0 K^{*-}_{0(2)}(1430) (e,f,g,h)$, and the annihilation one in the process $B^-\to \bar D^0 K^{*-}_{0(2)}(1430)(i,j,k,l)$. In the middle four diagrams, the exchange of $c$ and $u$ quark results in the corresponding diagrams for the process $B^-\to\bar D^0 K^{*-}_{0(2)}(1430)$.

Factorization formulas for  the $K^*_0$-emission diagrams  are given as
\begin{eqnarray}
 &&\xi_{ex}=N_1  f_{K^*_0} \int_{0}^{1}d x_{1}d
 x_{3}\int_{0}^{\infty} b_1d b_1 b_3d b_3
 \phi_B(x_1,b_1)
 \nonumber \\
 &&\times \phi_{D}(\bar x_3, b_3)\big[(2-x_3+r_D(2x_3-1)) E_a(t_a) a_1(t_a) h_a   \nonumber\\
 &&\;\; +r_D(1+r_D)E_b(t_b) a_1(t_b) h_b\big],
\end{eqnarray}
with $N_1=8\pi C_Fm_B^4$.  The hard kernels $E_i(t_i)$ and $h_{i}$  in these formulas are determined by the virtualities of the intermediate quarks and gluons and they can be found in the appendix~\ref{sec:hardkernels}.
The nonfactorizable contributions, the last diagrams in Fig.~\ref{fig:colorallowed},  have the formulas
\begin{eqnarray}
 &&{\cal M}_{ex}= N_2\int_0^1
 [dx]\int_0^{\infty} b_1db_1 b_2db_2
 \phi_B(x_1,b_1)\phi_{D}(\bar x_3,b_3)
 \nonumber \\
 & &\times\phi_{K^*_0}(x_2)
 \left[(x_2-r_D\bar x_3)E_c(t_c) h_{c}C_1(t_c)\right.\nonumber\\
 &&\left.+\big(\bar x_3r_{D}-\bar x_3-\bar x_2\big)E_{d}(t_d) h_{d} C_1(t_d)\right], \nonumber
\end{eqnarray}
with $\bar x_i=1-x_i$, $N_2=32\pi  m_B^4 {C_F}/{\sqrt{2N_c}}$, $[dx]\equiv dx_1dx_2dx_3$.
In the collinear approximation, the amplitude is divergent when the momentum fraction of the light spectator in the final state goes to zero.
The transverse momentum regulates this endpoint singularity and the threshold resummation function $S_t$ will suppress the endpoint contribution further.

The factorizable color-suppressed diagrams,  either the $D^0$ or $\bar D^0$ emission diagrams, have the same factorization formulas:
  \begin{eqnarray}
 &&\xi_{in}=\xi_{in}'=N_1f_{D}\int^1_0dx_1dx_3\int^\infty_0b_1db_1 b_3 db_3
\phi_{B}(x_1,b_1)
  \nonumber\\
  &&
  \times \Big\{ E_e(t_e)h_e a_2(t_e) \Big[ r_{K^*_0}(1-2x_3)(\phi_{K^*_0}^s(x_3)-\phi_{K^*_0}^T(x_3)) \nonumber\\
&&   +(2-x_3)\phi_{K^*_0}(x_3)
  \Big]    -2r_{K^*_0}\phi_{K^*_0}^s(x_3) E_f(t_f)h_fa_2(t_f)
 \Big\},
  \end{eqnarray}
but the $D^0$-emission nonfactorizable diagram is
\begin{eqnarray}
&& {\cal M}_{in} =N_2\int^1_0[dx]\int^\infty_0b_1db_1b_2db_2
\phi_{B}(x_1,b_1)\phi_D(\bar x_2,b_2)
\nonumber\\
&&\times
\Big\{\Big[x_2\phi_{K^*_0}(x_3)+r_{K^*_0}\bar x_3(\phi_{K^*_0}^s(x_3)+\phi_{K^*_0}^T(x_3))\Big] \nonumber\\
&&  \times  h_g E_g(t_g) C_1(t_h)-
  h_hE_h(t_h)C_2(t_h)
\nonumber\\
&&  \times\Big[(\bar x_2+\bar x_3)\phi_{K^*_0}(x_3)+r_3 \bar x_3(\phi_{K^*_0}^s(x_3)-\phi_{K^*_0}^T(x_3))\Big] \Big\}, \nonumber
 \end{eqnarray}
while the $\bar D^0$-emission is factorized as
\begin{eqnarray}
 &&{\cal M}_{in}' =N_2\int^1_0[dx]\int^\infty_0b_1db_1b_2db_2
\phi_{B}(x_1,b_1)\phi_D( x_2,b_2)
\nonumber\\
&&\times
\Big\{\Big[x_2\phi_{K^*_0}(x_3)+r_{K^*_0}\bar x_3 (\phi_{K^*_0}^s(x_3)+\phi_{K^*_0}^T(x_3))\Big]  \nonumber\\
 && \times h_g' E_g'(t_g')C_2(t_g')-h_h'
   E_h'(t_h^\prime)C_2(t_h') \nonumber\\
 && \times \Big[(\bar x_2+\bar x_3)\phi_{K^*_0}(x_3)+r_3\bar x_3(\phi_{K^*_0}^s(x_3)-\phi_{K^*_0}^T(x_3)\Big]  \Big\}.
 \end{eqnarray}
For the $B\to \bar D K_0^*$ decays, there are contributions from the
annihilation diagrams, which are depicted in
Fig.\ref{fig:colorallowed} $(i)(j)(k)(l)$. The amplitude of
factorizable annihilation diagrams are given as
\begin{eqnarray}
 &&\xi_{exc}=N_1f_{B}\int^1_0dx_2dx_3\int^\infty_0b_2db_2 b_3
 db_3\phi_D(x_3,b_3)
  \nonumber\\
  &&
  \times \Big\{ E_k(t_k)h_k a_1(t_k) \Big[ 2(x_3+1)r_D r_{K^*_0}\phi_{K^*_0}^s(x_2)\nonumber\\
  && -x_3\phi_{K^*_0}(x_2))\Big] + E_l(t_b)h_la_1(t_l)\nonumber\\
  &&\times\Big[r_D r_{K^*_0}((2x_2-3)\phi_{K^*_0}^s(x_2)-(2x_2-1)\phi_{K^*_0}^T(x_2))\nonumber\\
  &&-(x_2-1)\phi_{K^*_0}(x_2)\Big]
 \Big\}.
\end{eqnarray}
The nonfactorizable annihilation amplitude is given by
\begin{eqnarray}
&& {\cal M}_{exc} =N_2\int^1_0[dx]\int^\infty_0b_1db_1b_2db_2
\phi_{B}(x_1,b_1)\phi_D(x_3,b_3)
\nonumber\\
&&\times \Big\{\Big[r_D
r_{K^*_0}\big((x_2+x_3-1)\phi_{K^*_0}^T(x_2)-(x_2-x_3-3)\phi_{K^*_0}^s(x_2)\big)
\nonumber\\
&&+(x_2-1)\phi_{K^*_0}(x_2)\Big] h_m E_m(t_m) C_1(t_m)+
  h_nE_n(t_n)C_1(t_n)
\nonumber\\
&&  \times\Big[r_D
r_{K^*_0}\big((x_2-x_3-1)\phi_{K^*_0}^s(x_2)+(x_2+x_3-1)\phi_{K^*_0}^T(x_2)\big)\nonumber\\
&&+x_3\phi_{K^*_0}(x_2)\Big] \Big\}, \nonumber
\end{eqnarray}

The formulas for channels involving $K^*_2$ are obtained by the replacement  $f_{K^*_0}\to 0$, $\phi_{K^*_0}\to \phi_{K^*_2}$ and $\phi_{K^*_0}^{s,T}\to \phi_{K^*_2}^{s,t}$. Incorporating the CKM matrix elements, we have the total decay amplitudes
\begin{eqnarray}
 A(B^-\to \bar D^0 K^{*-}_{0,2})  &&=\frac{G_F}{\sqrt 2} V_{ub}V_{cs}^* \nonumber\\
  &&\times \left( \xi_{in}'+ {\cal M}_{in}' + \xi_{exc} + {\cal M}_{exc} \right), \nonumber \\
 A(B^-\to  D^0 K^{*-}_{0,2})  &&=\frac{G_F}{\sqrt 2}  V_{cb}V_{us}^* \nonumber\\
&& \times \big(\xi_{ex} + {\cal M}_{ex}  + \xi_{in}+ {\cal M}_{in}\big),\label{eq:factorization}
\end{eqnarray}
where
$G_F$ is the Fermi constant. It should be pointed out that the  color-allowed  $\xi_{ex}$ is zero in $B^-\to D^0 K^{*-}_2$ due to the fact that the tensor meson can not be generated by a local vector or axial-vector current.

\section{Numerical Results and Discussions} \label{sec:f0}

The expression for $\phi_{B}(x,b)$ has been examined in various kinds of $B$ decays and the currently-accepted form in the PQCD approach is
 \begin{equation}
 \phi_{B}(x,b)=N_{B}x^2(1-x)^2\mbox{exp}\Big[-\frac{m_{B}^2 x^2}{2\omega_b^2}
-\frac{1}{2}(\omega_b  b)^2\Big], \label{Bwave:da}
 \end{equation}
where  the normalization factor $N_{B}$ is related to the decay constant $f_B$. We adopt the ansatz that the $B$ meson wave functions  have a sharp peak at $x\sim 0.1$, in accordance with the
most probable momentum fraction of the light quark.
The best-fitted form for $\phi_{D}$  from the $B$ meson decays into a charmed meson derived in Refs.~\cite{Keum:2003js,Li:2008ts,Zou:2009zza} is
\begin{eqnarray}
\phi_{D}(x_2,b_2)&=& \frac{f_D}{ 2\sqrt {2N_c}} 6x(1-x) [1+C_D (1-2x)]  \nonumber\\
&& \times
 {\rm exp}\left[- {\omega_D^2 b_2^2}/{2}\right].
\end{eqnarray}
Their numerical values (in   GeV except $C_D$) are used as
\begin{eqnarray}
&& C_{D}=(0.5\pm 0.1),\;\;  \omega_b=(0.40\pm 0.05), \;\; \omega_{D}=0.1, \nonumber\\
&& f_{B}=(0.1969\pm 0.0089),\;\;f_{D}= (0.221\pm 0.018),
\end{eqnarray}
where  $f_B$ is from the recent Lattice QCD simulation~\cite{Neil:2011ku} and the $f_D$ is extracted from $D^-\to \mu\bar\nu_\mu$~\cite{Nakamura:2010zzi}.

For the LCDAs of the light scalar meson $K^*_0$, we adopt $B_0= ({m_s-m_u})/{m_{K^*_0}} =0.07$~\cite{Nakamura:2010zzi} and  the two different solutions in Ref.~\cite{Cheng:2005nb}
\begin{eqnarray}
 S1:\;\;  \bar f_{K^*_0}= (-300\pm30) {\rm MeV} , \;\; B_1= 0.58\pm 0.07, \nonumber\\
 B_3= -1.20\pm 0.08,\nonumber\\
S2:\;\;    \bar  f_{K^*_0}=(445\pm50) {\rm MeV} , \;\; B_1=-0.57\pm 0.13, \nonumber\\ B_3=-0.42\pm 0.22.
\end{eqnarray}
The normalization constants in $K^*_2$ LCDAs are~\cite{Cheng:2010hn}
\begin{eqnarray}
 f_{K^*_2}=(118\pm 5) {\rm MeV},\;\; f_{K^*_2}^T= (77\pm 14) {\rm MeV}.
\end{eqnarray}
These LCDAs have been used to calculate the form factors of $B$
decays into a scalar/tensor meson in the same
perturbative QCD approach~\cite{Wang:2006ria,Wang:2010ni,Zou,Freddy:2013apa}.

For the CKM matrix elements, we use~\cite{Nakamura:2010zzi}
\begin{eqnarray}
 |V_{ub}| &=& (3.89\pm 0.44)\times 10^{-3},\;\; \nonumber\\
 |V_{cs}| &=& 0.97345,\nonumber\\
 |V_{us}| &=& 0.2252, \nonumber\\
 |V_{cb}| &=& (40.6\pm 1.3)\times 10^{-3},
\end{eqnarray}
where the small uncertainties are not taken into account.

 With the above inputs, we predict the branching ratios as
\begin{eqnarray}
 {\cal B}(B^-\to D^0 K^{*-}_{0}) &=& \left(2.70_{-0.97-0.48}^{+1.09+0.25}\right) \times 10^{-4}, \;\;\; S1\nonumber\\
 {\cal B}(B^-\to \bar  D^0 K^{*-}_{0}) &=& \left(1.53_{-0.53-0.46}^{+0.82+0.62}\right) \times 10^{-5},\;\;\; S1\nonumber\\
 {\cal B}(B^-\to D^0 K^{*-}_{0}) &=& \left(1.16_{-0.41-0.22}^{+0.40+0.14}\right) \times 10^{-4}, \;\;\; S2\nonumber\\
 {\cal B}(B^-\to \bar  D^0 K^{*-}_{0}) &=& \left(3.38_{-1.18-0.59}^{+1.51+0.65}\right) \times 10^{-5},\;\;\; S2\nonumber\\
 {\cal B}(B^-\to D^0 K^{*-}_{2}) &=& \left(2.40_{-0.97-1.05}^{+1.30+0.72}\right) \times 10^{-5},\nonumber\\
 {\cal B}(B^-\to \bar  D^0 K^{*-}_{2}) &=& \left(3.32_{-1.18-0.74}^{+1.90+1.07}\right) \times 10^{-6},
\end{eqnarray}
where the first uncertainties are from $f_B$ and $\omega_b$ in the $B$ meson wave functions, the second errors are from $\Lambda_{QCD}$ and the scales defined in Appendix \ref{sec:hardkernels} ( We vary the $\sqrt{Q}$ and $\sqrt{P}$ in the scales $25\%$ for error estimation).
The results for the branching ratios made here are   larger than our previous estimates in Ref.~\cite{Wang:2011zw},  obtained under the factorization approach. The main reason is due to the enhancement of nonfactorizable
contributions in color-suppressed $D^0$-emission.
Ratios and phases of the amplitudes are
\begin{eqnarray}
 r_{K^*_0} &=& 0.24_{-0.01-0.04}^{+0.02+0.07}, \;\; \delta_{K^*_0}=\left(-125.65_{-0.00-17.42}^{+3.95+23.16}\right)^\circ,\;\;\; S1\nonumber\\
 r_{K^*_0} &=& 0.54_{-0.02-0.07}^{+0.03+0.11}, \;\; \delta_{K^*_0}=\left(-161.51_{-0.16-9.53}^{+0.91+12.01}\right)^\circ,\;\;\; S2\nonumber\\
 r_{K^*_2} &=& 0.37_{-0.00-0.09}^{+0.02+0.17}, \;\; \delta_{K^*_2}=\left(155.53_{-3.36-3.49}^{+0.00+2.98}\right)^\circ.
\end{eqnarray}
It should be pointed out that although the large uncertainties in many entries like decay constants will affect our predictions for branching ratios, the relative strength of decay amplitudes are almost unaffected.
%  In fact  in Ref.~\cite{Wang:2011zw} we have pointed out that using the instance of $B^-\to DK^\pm$  the factorization approach will undershoot the experimental data.

\begin{figure}[h]
\begin{center}
\includegraphics[scale=0.35]{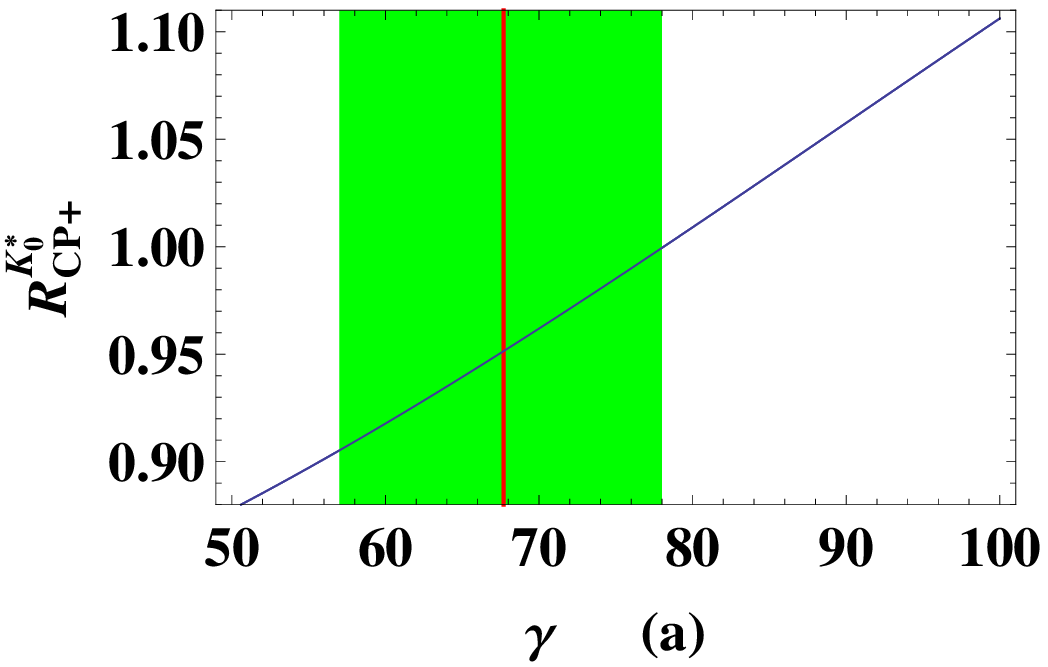}
\includegraphics[scale=0.38]{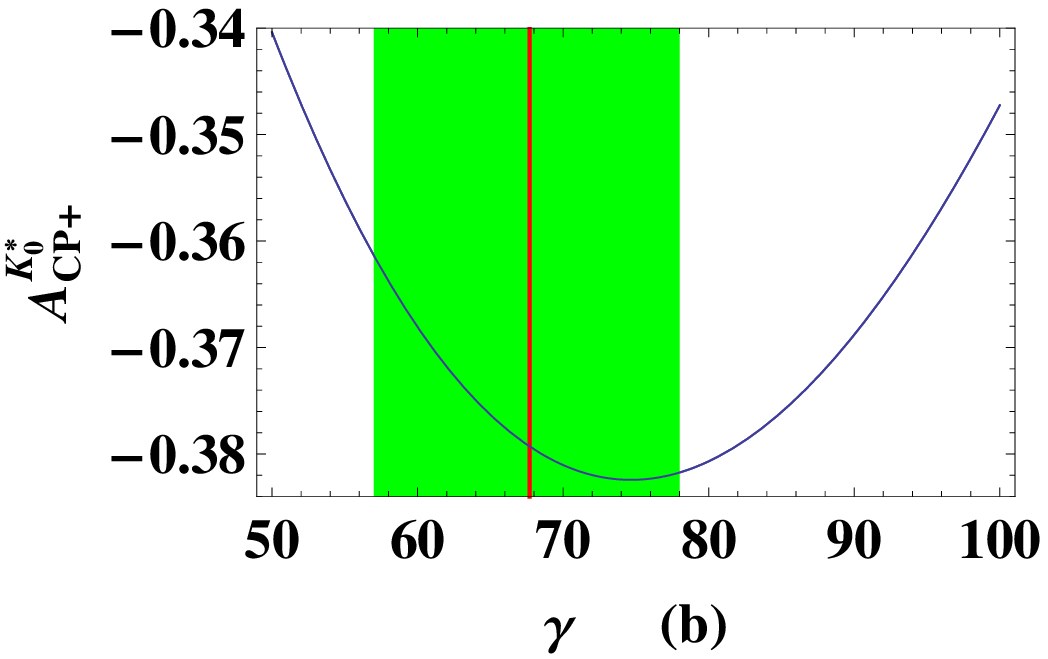}
\includegraphics[scale=0.35]{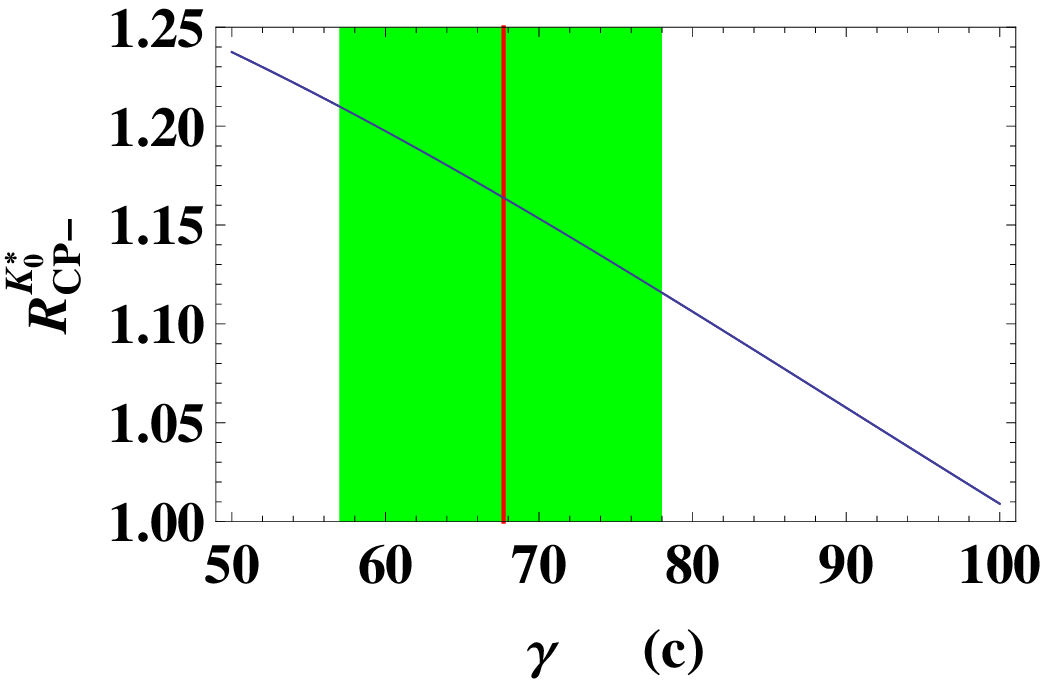}
\includegraphics[scale=0.38]{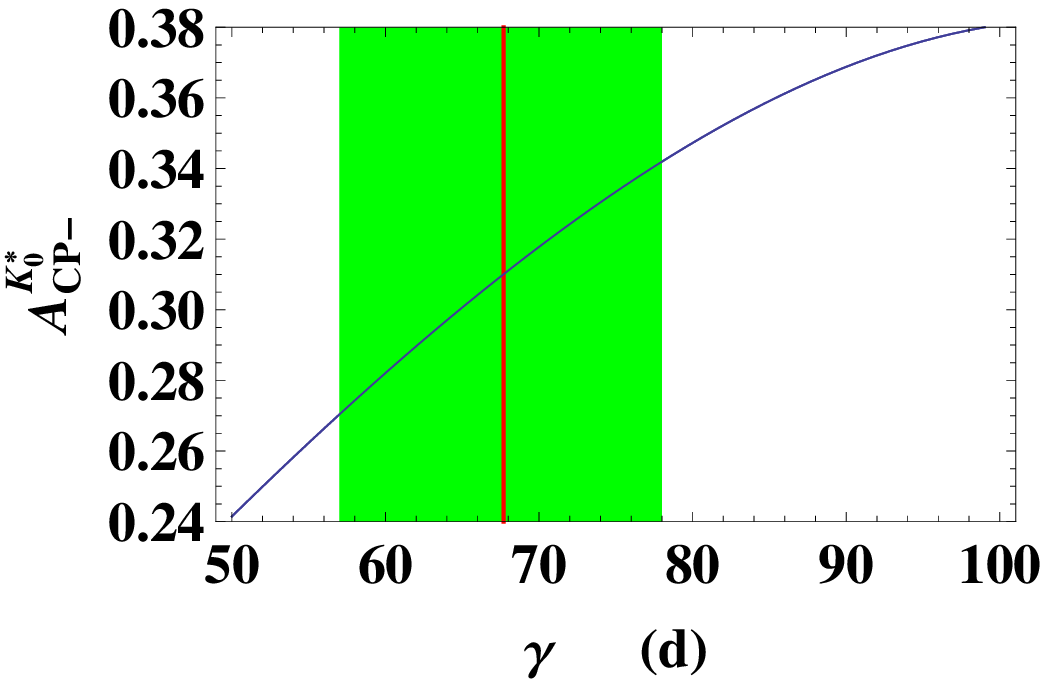}
\includegraphics[scale=0.35]{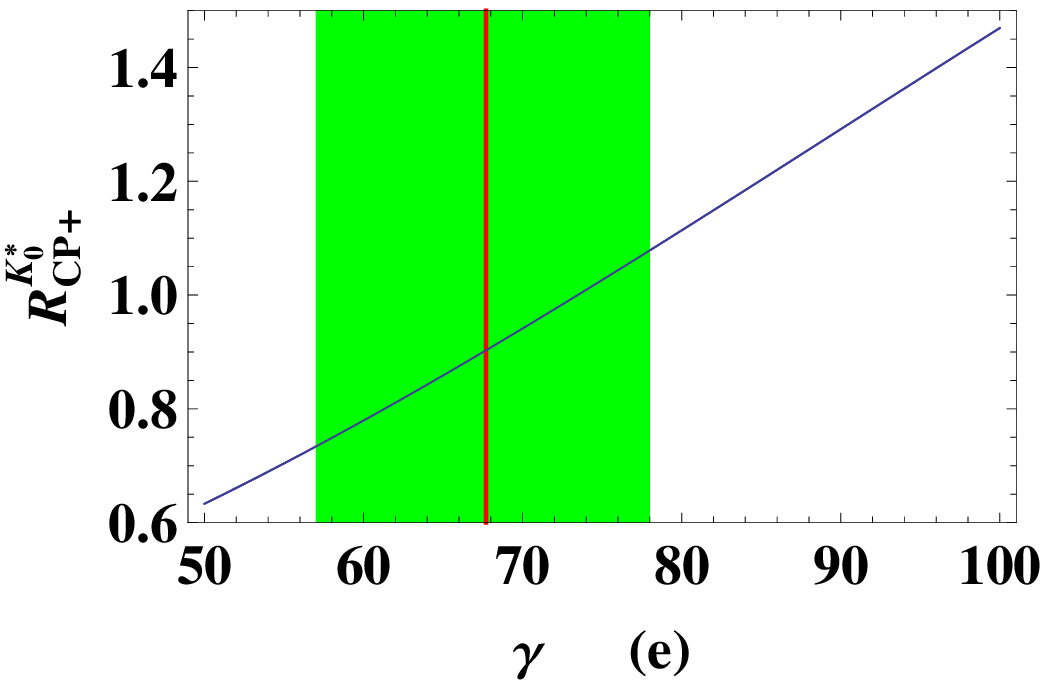}
\includegraphics[scale=0.38]{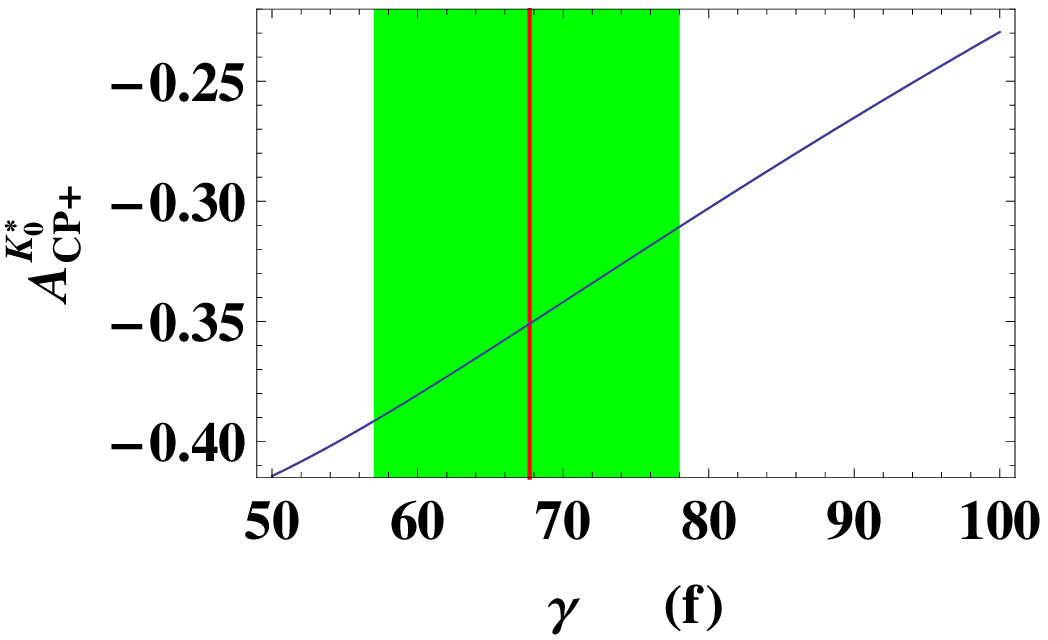}
\includegraphics[scale=0.35]{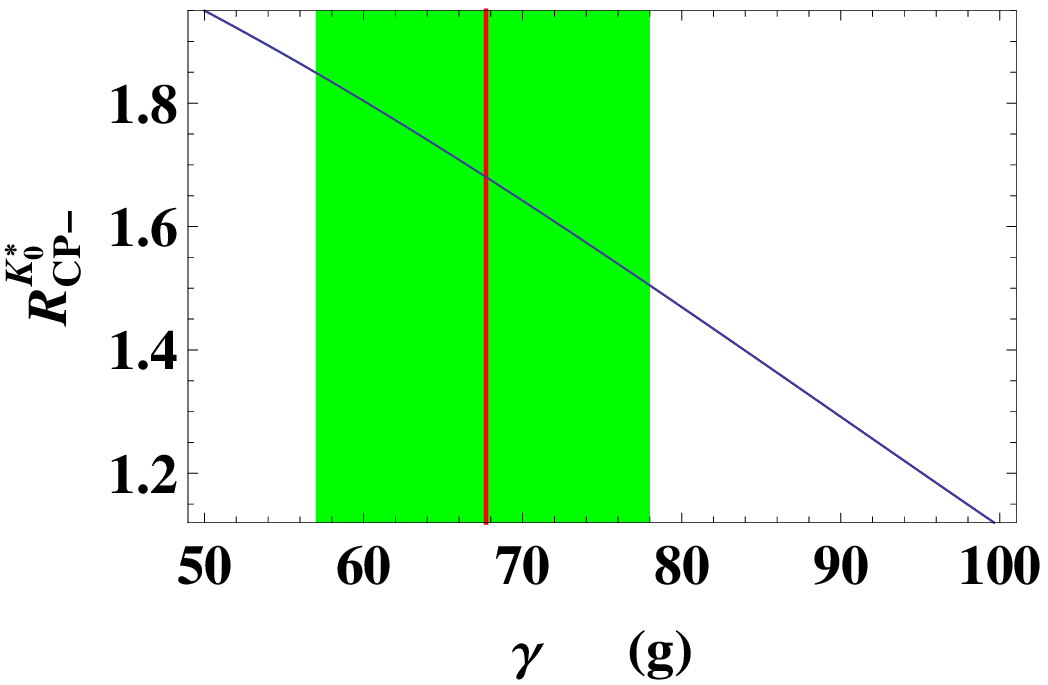}
\includegraphics[scale=0.38]{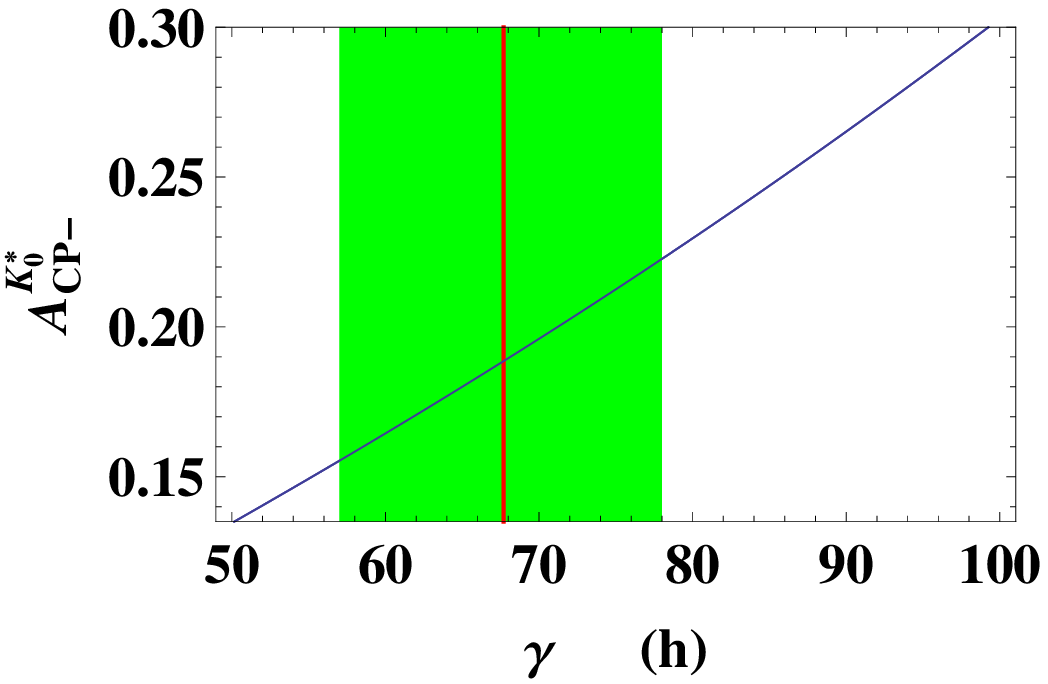}
\includegraphics[scale=0.35]{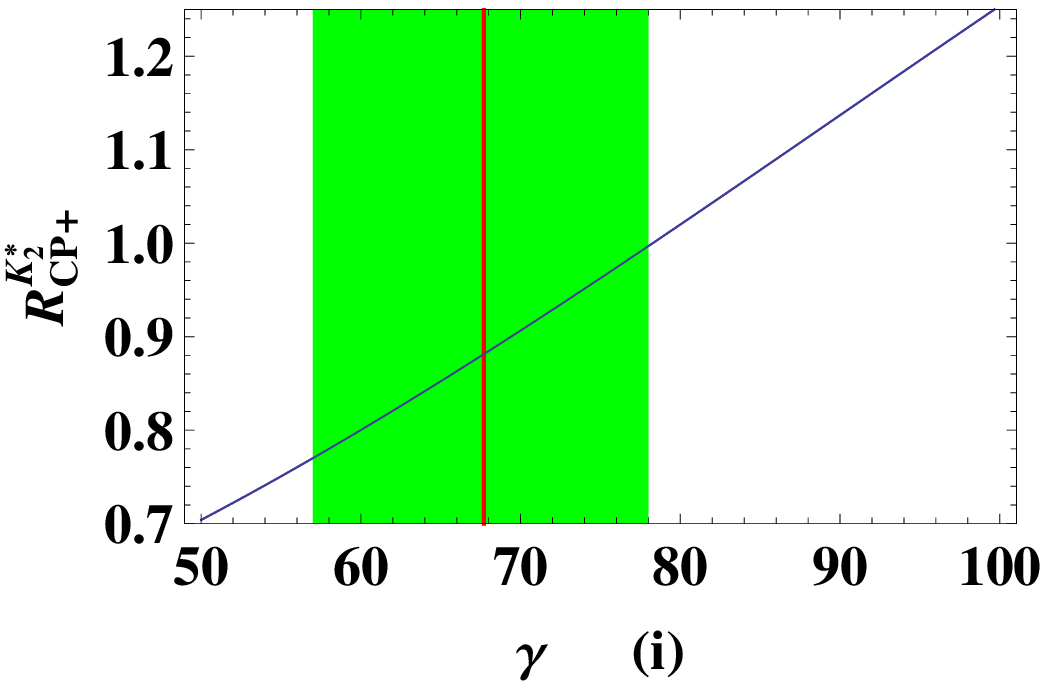}
\includegraphics[scale=0.38]{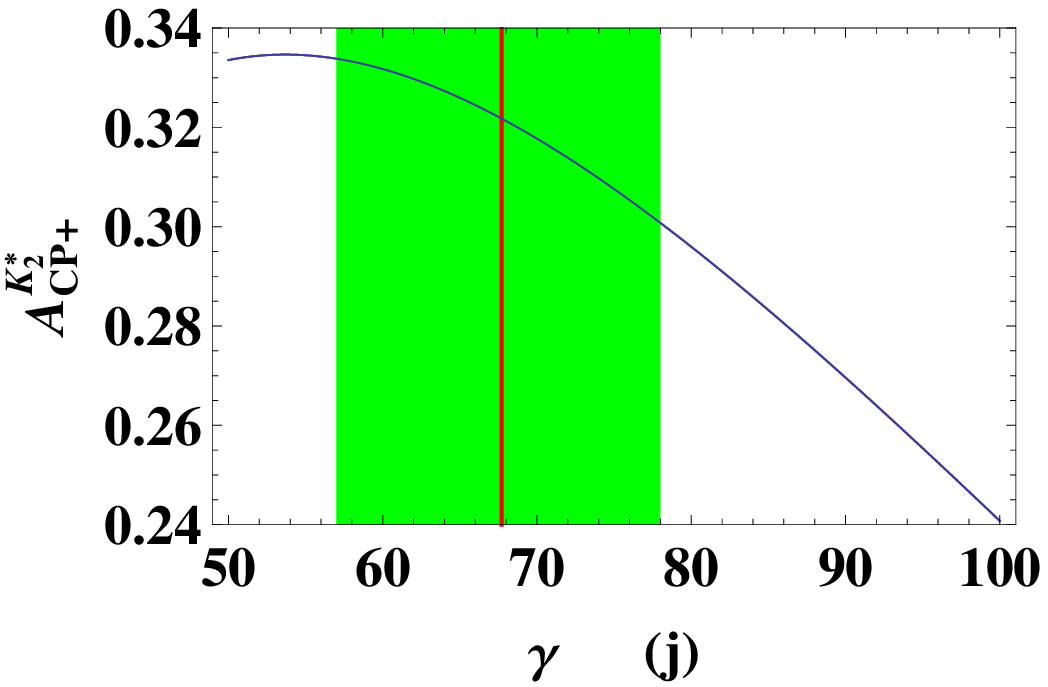}
\includegraphics[scale=0.35]{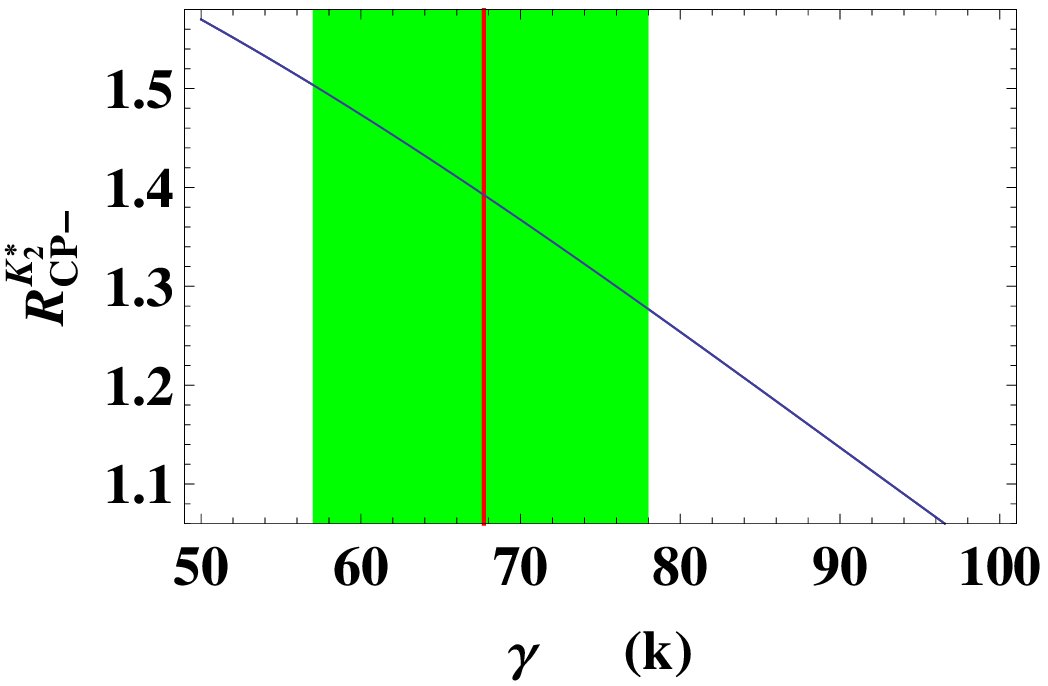}
\includegraphics[scale=0.38]{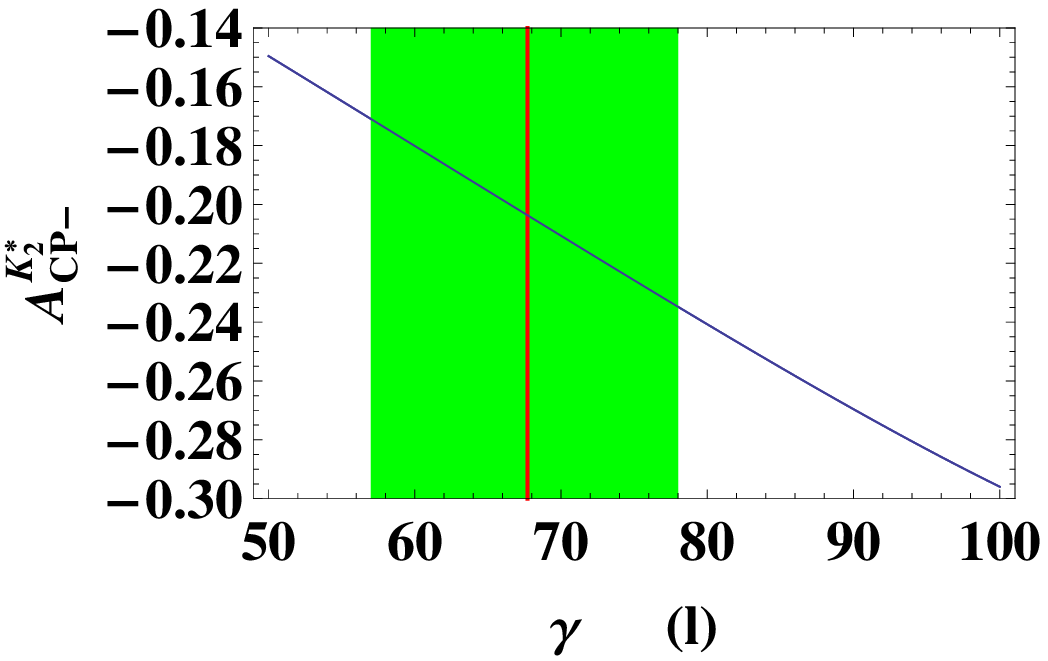}
\caption{The dependence of $R_{CP}$ and $A_{CP}$ on $\gamma$. Diagrams (a)-(d) show $R^{K_0^*}_{CP}$ and $A^{K_0^*}_{CP}$ in $S1$, (e)-(h) in $S2$, and diagrams (i)-(l) show $R^{K_2^*}_{CP}$ and $A^{K_2^*}_{CP}$. The shadowed (green) region denotes the current bounds on $\gamma=(68_{-11}^{+10})^\circ$ from a combined analysis of $B^\pm\to DK^\pm$ \cite{CKMfitter}, and the vertical (red) line represents the central value. }
\label{fig:dependence}
\end{center}
\end{figure}

 Physical observables that are experimentally explored are defined as
\begin{eqnarray}
 R_{CP\pm}^{K_J} &=&2\frac{{\cal B}(B^-\to D_{CP\pm} {K_J}^-)+{\cal B}(B^+\to D_{CP\pm} K_J^+)  }{{\cal B}(B^-\to D^0K_J^-) +{\cal B}(B^+\to \bar D^0 K_J^+) }\nonumber\\
&=& 1+(r_{B}^{K_J})^2\pm 2r_{B}^{K_J} \cos\delta_{B}^{K_J} \cos\gamma,\nonumber\\
 A_{CP\pm}^{K_J} &=&\frac{{\cal B}(B^-\to D_{CP\pm} {K_J}^-)-{\cal B}(B^+\to D_{CP\pm} K_J^+)  }{{\cal B}(B^-\to D_{CP\pm} K_J^-) +{\cal B}(B^+\to D_{CP\pm} K_J^+) }\nonumber\\
&=&\pm 2r_B^{K_J} \sin\delta_{B}^{K_J} \sin\gamma /R_{CP\pm}^K.
\end{eqnarray}
In the limit of $r_{B}\to 0$, the ratio $R_{CP\pm}^K$ is close to 1 while the CP asymmetries vanish.
As we have  pointed out, due to the suppression of the color-allowed decay amplitudes based on the fact that  the matrix element of a local vector or  axial-vector current (at the lowest order in $\alpha_s$) between the QCD vacuum and  the $K^*_{0}$($K^*_{2}$) state is small (identically zero) , the low sensitivity to $\gamma$ is  improved and in particular large CP asymmetries are expected. The dependence of  $R_{CP}$ and $A_{CP}$ on $\gamma$ is shown in Fig.~\ref{fig:dependence}. Since the errors of $r_{K_{0,2}^*}$ and $\delta_{K_{0,2}^*}$ are not large, only their central values are used. We investigate these observables in the region $\gamma=(68_{-11}^{+10})^\circ$ which is from a combined analysis of $B^\pm\to DK^\pm$ \cite{CKMfitter}. In this region we find that the observables of the $B^-\to(\bar D^0,D^0)K_0^{*-}$ in $S1$ have relative smaller variances because of the smaller $r_{K^*_0}$, most of which are around $10\%$. However, for the other cases the observables have large variances, and some of them even reach about $40\%$. Therefore these channels have the potential to improve the accuracy of $\gamma$ extracted from the $B^\pm\to DK^\pm$ decays.

It is also interesting to notice that due to the large value of $r_{K_{0,2}^*}$, the large impact arising from the direct CP violation  of $D^0$ decays into CP eigenstates $K^+K^-/\pi^+\pi^-$, of the order ${\cal O}(A_{CP}^{dir})/r_{K_{0,2}^*}$ ~\cite{Wang:2012ie,Martone:2012nj,Bhattacharya:2013vc} are not important in $B\to DK_{0,2}^*$.

%%%%%%%%%%%%%%%%%%%%%%%%%%%%%%%%%%%%%%%%%%%%%%%%%%%%
%%%%%%%%%%%%%%%%%%%%%%%%%%%%%%%%%%%%%%%%%%%%%%%%%%%%%

 As discussed in Ref.~\cite{Wang:2011zw}, the time-dependent observables of $B_s\to (D,\bar D)f_0(980)$ and $B_s\to (D,\bar D)f_2'(1525)$ processes can be used to determine the $\gamma$ as well. Therefore we will also predict their branching ratios in the perturbative QCD approach. In these channels only the color-suppressed diagrams depicted as Fig. \ref{fig:colorallowed}$(e)(f)(g)(h)$ contribute, in which the spectator quark $\bar u$ need to be replaced by $\bar s$ and the $c$ and $\bar u$ in the emission meson should be exchanged for the $B_s\to\bar D (f_0,f_2^{\prime})$ decays. The amplitudes are given by
 \begin{eqnarray}
  A(\bar B_s^0\to \bar D^0 (f_0,f_2^\prime))  &&=\frac{G_F}{\sqrt 2} V_{ub}V_{cs}^*  \left( \xi_{in}'+ {\cal M}_{in}' \right), \nonumber \\
 A(\bar B_s^0\to D^0 (f_0,f_2^\prime))  &&=\frac{G_F}{\sqrt 2}  V_{cb}V_{us}^* \left( \xi_{in}+ {\cal M}_{in}\right).\label{eq:BtoDf}
 \end{eqnarray}
Our inputs for the $\bar B_s^0\to (\bar D,D)(f_0,f_2^\prime)$ decays are summarized as (decay constants in units of GeV)~\cite{Cheng:2005nb,Cheng:2010hn,Neil:2011ku}
\begin{eqnarray}
 f_{B_s}= 0.2420\pm 0.0095,\;\; \bar  f_{f_0}=0.37\pm0.02,\nonumber\\
  B_1(f_0)=-0.78\pm 0.08, \;\; B_3(f_0)=0.02\pm 0.07 ,\nonumber\\
 f_{f_2'}=0.126\pm 0.004,\ f_{f_2'}^T= 0.065\pm 0.012,
\end{eqnarray}
with which the branching ratios are predicted as
\begin{eqnarray}
 {\cal B}(\bar B_s^0\to D^0 f_0) &=& \left(3.50_{-1.15-0.77}^{+1.26+0.56}\right)\times 10^{-5},\nonumber\\
 {\cal B}(\bar B_s^0\to \bar D^0 f_0) &=& \left(5.94_{-2.16-0.97}^{+3.13+1.47}\right)\times 10^{-6},\nonumber\\
 {\cal B}(\bar B_s^0\to D^0 f_2') &=&\left(1.08_{-0.35-0.28}^{+0.37+0.26}\right)\times 10^{-5} ,\nonumber\\
 {\cal B}(\bar B_s^0\to \bar D^0 f_2') &=& \left(2.85_{-0.95-0.43}^{+1.53+0.67}\right)\times 10^{-6}.
\end{eqnarray}

\section{Summary}\label{sec:conclusions}

The determination of the CKM angles is crucial for the test of the CKM paradigm and also sheds light on the standard model description of the CP violation.  To accomplish this goal, one of the most important efforts to be done in the next step is to reduce the uncertainties in these entries. What  has been explored in Ref.~\cite{Wang:2011zw} and this work is to propose that the $B\to D K_{0,2}^*$ is expedient to  provide complementary  information of the angle $\gamma$.

In this work we have calculated the branching ratios of $B\to D K_{0,2}^*$ and the corresponding $B_s$ relatives, by adopting the $k_T$ factorization approach.  We find that the BR of  $B\to D K_{0(2)}^*$ can reach $10^{-4}(10^{-5})$ while the ratio of the magnitude is also  significantly enhanced compared to $B\to DK$ mode. As a consequence, it seems promising for the LHCb experiment and the currently-designed Super B factory to measure  $B_{u,d}\to DK^{*}_{0(2)}(1430)$ and  time-dependent CP asymmetries in the  $B_s$ decays.

\section*{Acknowledgement}
The work of CSK and RHL is supported by the National Research Foundation
of Korea (NRF) grant funded by Korea government of the Ministry of Education, Science and Technology (MEST) (Grant No. 2011-0017430 and Grant No. 2011-0020333).
The work of WW is  supported in
part by the DFG and the NSFC through funds provided to
the Sino-German CRC 110 ``Symmetries and the Emergence
of Structure in QCD". The computation code can be found: http://www.itkp.uni-bonn.de/$\sim$weiwang/.

\begin{appendix}

\section{Hard kernels in the PQCD calculation}\label{sec:hardkernels}

The offshellness of the intermediate gluon
\begin{eqnarray}
 &&Q_{a,b,c,d}= x_1\bar x_3 m_{B}^2,\;\;   Q_{e,f,g,h,g',h'}= x_1 \bar x_3(1-r_D^2) m_{B}^2,
 \nonumber\\
 &&Q_{k,l,m,n}= x_3\bar x_2(1-r_D^2) m_B^2.\nonumber
\end{eqnarray}
and the quarks
\begin{eqnarray}
 &&P_{a}= \bar x_3 m_{B}^2 ,\;
 P_{b}= x_1 m_{B}^2 ,\; \nonumber\\
 &&P_{c}= \bar x_3 [x_1-x_2(1-r_{D}^2)] m_{B}^2,\nonumber\\
 &&P_{d}= \bar x_3 [x_1-\bar x_2(1-r_{D}^2)] m_{B}^2,\nonumber\\
 &&P_{e}= \bar x_3 (1-r_{D}^2)m_{B}^2 ,\;
 P_{f}= x_1(1-r_{D}^2) m_{B}^2 ,\; \nonumber\\
 &&P_{g}= \left\{(x_1-x_2) [(1-r_D^2) \bar x_3 +r_D^2] +r_D^2\right\}m_{B}^2,\nonumber\\
 &&P_{h}= \left\{(x_1-\bar x_2) (1-r_D^2) \bar x_3 \right\}m_{B}^2,\nonumber\\
 &&P_{g}'= \left\{(x_1-x_2) (1-r_D^2) \bar x_3 \right\}m_{B}^2,\nonumber\\
 &&P_{h}= \left\{(x_1-\bar x_2) [(1-r_D^2) \bar x_3+r_D^2]+r_D^2 \right\}m_{B}^2,\nonumber\\
 &&P_{k}= x_3(1-r_D^2)m_B^2,\;
 P_{l}= \bar x_2(1-r_D^2)m_B^2,\nonumber\\
 &&P_{m}= -\left\{\bar x_3\left[1-(1-r_D^2)\bar x_2 -x_1\right]-1\right\}m_B^2,\nonumber\\
 &&P_{n}= x_3\left[x_1-\bar x_2(1-r_D^2)\right]m_B^2,\nonumber
\end{eqnarray}
result in the  hard scales and kernels
\begin{eqnarray}
t_i&=&\mbox{max}\{\sqrt{Q_{i}}, \sqrt{P_{i}} ,1/b_1,1/b_3\},\nonumber\\
t_j&=&\mbox{max}\{\sqrt{Q_j}, \sqrt{P_{j}} ,1/b_1,1/b_2\},\nonumber\\
t_x&=&\mbox{max}\{\sqrt{Q_{x}}, \sqrt{P_{x}} ,1/b_2,1/b_3\},\nonumber\\
t_y&=&\mbox{max}\{\sqrt{Q_y}, \sqrt{P_{y}} ,1/b_1,1/b_2\},\nonumber\\
 h_{a,e}&=& H_{e}( P_{a,e}, Q_{a,e}, b_1, b_3) S_{t}(x_3),\nonumber\\
 h_{b,f}&=& H_{e}( P_{b,f}, Q_{b,f}, b_3, b_1)S_t(x_1),\nonumber\\
 h_{j}&=& H_{en}( Q_{j}, P_{j}, b_1,b_2),\nonumber\\
 h_k&=& H_{af}( P_k, Q_k, b_2, b_3) S_{t}(x_3),\nonumber\\
 h_l&=& H_{af}( P_l, Q_l, b_3, b_2) S_{t}(x_2),\nonumber\\
 h_{m,n}&=& H_{an},
\end{eqnarray}
for the factorizable diagram $i=a,b, e,f$, for the nonfactorizable
diagram $j=c,d,g,h,g',h'$, and for the annihilation diagrams $x=k,l$
and $y=m,n$. For diagrams (c,d), we also keep the function
$S_t(x_3)$. Here we use the definition of Bessel functions
\begin{eqnarray}
&&H_e(\alpha, \beta, b_1, b_3)=K_0(\sqrt {\beta} b_1)   \big[\theta(b_1-b_3)I_0(\sqrt
\alpha b_3) \nonumber\\
&& \times K_0(\sqrt
\alpha  b_1)  +(b_1\leftrightarrow b_3)\big], \nonumber\\
&&H_{en}(\alpha, \beta,b_1,b_2)=\left[\theta(b_2-b_1)K_0(\sqrt
{\alpha} b_2) \times  I_0(\sqrt
{\alpha} b_1) \right. \nonumber\\
&& \left.+(b_1\leftrightarrow b_2)\right]
\left\{\begin{array}{ll}\frac{i\pi}{2}H_0^{(1)}(\sqrt{\beta}
 b_2),&\beta<0\\
 K_0(\sqrt{\beta}b_2),& \beta >0
\end{array}
\right.
\end{eqnarray}
\begin{eqnarray}
 %---------------------------------------------------------------------
&&H_{af}(\alpha,\beta,b_2,b_3)=\left(i\frac{\pi}{2}\right)^2
H_0^{(1)}\left(\sqrt{\beta}b_2\right)
\nonumber \\
& &\times\left[\theta(b_2-b_3)
H_0^{(1)}\left(\sqrt{\alpha}b_2\right)
J_0\left(\sqrt{\alpha}b_3\right)\right.
\nonumber \\
& &\left.+\theta(b_3-b_2)H_0^{(1)}\left(\sqrt{\alpha}b_3\right)
J_0\left(\sqrt{\alpha}b_2\right)\right]\;,\nonumber\\
%--------------------------------------------------------------------------------
&&H_{an}= i\frac{\pi}{2}
\left[\theta(b_1-b_2)H_0^{(1)}\left(\sqrt{Q_y}
b_1\right)J_0\left(\sqrt{Q_y}b_2\right)\right.\nonumber\\
&&\left.+\theta(b_2-b_1)H_0^{(1)}\left(\sqrt{Q_y} b_2\right)
J_0\left(\sqrt{Q_y} b_1\right)\right]\;  \nonumber \\
&  & \times \left\{ \begin{array}{cc}
K_{0}(\sqrt{|P_y|}b_{1}) &  \mbox{for $P_y \leq 0$}  \\
\frac{i\pi}{2} H_{0}^{(1)}(\sqrt{|P_y}b_{1})  & \mbox{for $P_y \geq
0$}
\end{array} \right\}\;,
\end{eqnarray}
where $H_0^{(1)}(z) = \mathrm{J}_0(z) + i\, \mathrm{Y}_0(z)$. The
$S_t$ has been parameterized as
  \begin{eqnarray}
S_t(x)=\frac{2^{1+2c}\Gamma(3/2+c)}{\sqrt \pi
\Gamma(1+c)}[x(1-x)]^c,
\end{eqnarray}
with $c=0.4\pm 0.1$.

The  $E_i$s  contain the Sudakov evolution factors and the strong coupling constants, and  are
given by
\begin{eqnarray}
E_{a,b}(t)&=&\alpha_s(t)  \exp[-S_B(t)-S_D(t,\bar x_3, b_3)],
 \nonumber \\
 E_{c,d}(t)&=&\alpha_s(t)
 \exp[-S_B(t)-S_D(t,\bar x_3, b_1)-S_{K^*_0}(t, b_2)],\nonumber\\
E_{e,f}(t)&=&\alpha_s(t)   \exp[-S_B(t)-S_{K^*_0}(t, b_3)],\nonumber\\
 E_{g,h}(t)&=& \alpha_s(t)
 \exp[-S_B(t)-S_D(t, \bar x_2, b_2)-S_{K^*_0}(t, b_1)],\nonumber\\
 E_{g',h'}(t)&=& \alpha_s(t)
 \exp[-S_B(t)-S_D(t, x_2, b_2)-S_{K^*_0}(t,b_1)],\nonumber\\
  E_{k,l}(t)&=& \alpha_s(t)
 \exp[-S_D(t, x_3, b_3)-S_{K^*_0}(t, b_2)],\nonumber\\
 E_{m,n}(t)&=& \alpha_s(t)
 \exp[-S_B(t)-S_D(t, x_2, b_2)-S_{K^*_0}(t, b_2)].\nonumber
\end{eqnarray}
in which the Sudakov exponents are defined as
\begin{eqnarray}
S_B(t)&=&s\left(x_1\frac{m_B}{\sqrt
2},b_1\right)+\frac{5}{3}\int^t_{1/b_1}\frac{d\bar \mu}{\bar
\mu}\gamma_q(\alpha_s(\bar \mu)),\nonumber\\
S_D(t,x_2, b_2)&=&s\left(x_2\frac{m_B}{\sqrt
2},b_2\right) +2\int^t_{1/b_2}\frac{d\bar \mu}{\bar
\mu}\gamma_q(\alpha_s(\bar \mu)),\nonumber\\
S_{K^*_0}(t,b_3)&=&s\left(x_3\frac{m_B}{\sqrt
2},b_3\right)+s\left((1-x_3)\frac{m_B}{\sqrt
2},b_3\right) \nonumber\\
&& +2\int^t_{1/b_3}\frac{d\bar \mu}{\bar
\mu}\gamma_q(\alpha_s(\bar \mu)),
\end{eqnarray}
 with the quark
anomalous dimension $\gamma_q=-\alpha_s/\pi$.

\end{appendix}

%%%%%%%%%%%%%%%%%%%%%%%%%%%%%%%%%%%%%%%%%%%%%%%%%%%%%%%%%%


\begin{thebibliography}{11}


%\cite{Asner:2010qj}
\bibitem{Asner:2010qj}
 D.~Asner {\it et al.} [HFAG ],
%  ``Averages of b-hadron, c-hadron, and $\tau-lepton Properties,''
   arXiv:1207.1158 [hep-ex]; updated results available at: http://www.slac.stanford.edu/xorg/hfag/.

   %\cite{Lees:2013zd}
   \bibitem{Lees:2013zd}
For recent measurements please see:
%
  J.~P.~Lees {\it et al.}  [BaBar Collaboration],
  %``Observation of direct CP violation in the measurement of the Cabibbo-Kobayashi-Maskawa angle gamma with $B^\pm\to D^{(*)}K^{(*)\pm}$ decays,''
  Phys.\ Rev.\ D {\bf 87}, 052015 (2013)
  [arXiv:1301.1029 [hep-ex]];
   LHCb Collaboration,  LHCb-CONF-2012-032;
%\bibitem{Trabelsi:2013uj}
  K.~Trabelsi [Belle Collaboration],
  %``Study of direct CP in charmed B decays and measurement of the CKM angle gamma at Belle,''
  arXiv:1301.2033 [hep-ex];
   %\cite{Ricciardi:2013mca}
%\bibitem{Ricciardi:2013mca}
  S.~Ricciardi [LHCb Collaboration],
  %``Measurements of the CKM angle gamma in tree-dominated decays at LHCb,''
  arXiv:1302.4582 [hep-ex].
  %%CITATION = ARXIV:1302.4582;%%
  %1 citations counted in INSPIRE as of 21 May 2013
  %\cite{Trabelsi:2013uj}
  %%CITATION = ARXIV:1301.2033;%%
  %5 citations counted in INSPIRE as of 21 May 2013  %%CITATION = ARXIV:1301.1029;%%
  %6 citations counted in INSPIRE as of 21 May 2013


 %\cite{Gronau:1991dp}
\bibitem{Gronau:1991dp}
  M.~Gronau, D.~Wyler,
  %``On determining a weak phase from CP asymmetries in charged B decays,''
  Phys.\ Lett.\  {\bf B265}, 172-176 (1991).

%\cite{Gronau:1990ra}
\bibitem{Gronau:1990ra}
  M.~Gronau, D.~London,
  %``How to determine all the angles of the unitarity triangle from B(d)0 ---> D K(s) and B(s)0 ---> D0,''
  Phys.\ Lett.\  {\bf B253}, 483 (1991).

%\cite{Dunietz:1991yd}
\bibitem{Dunietz:1991yd}
  I.~Dunietz,
  %``CP violation with selftagging B(d) modes,''
  Phys.\ Lett.\  {\bf B270}, 75-80 (1991).

%\cite{Atwood:1996ci}
%\bibitem{Atwood:1996ci}
 % D.~Atwood, I.~Dunietz, A.~Soni,
  %``Enhanced CP violation with B ---> K D0 (anti-D0) modes and extraction of the CKM angle gamma,''
  %Phys.\ Rev.\ Lett.\  {\bf 78}, 3257 (1997).
%  [hep-ph/9612433].

%\cite{Atwood:2000ck}
%\bibitem{Atwood:2000ck}
 % D.~Atwood, I.~Dunietz, A.~Soni,
  %``Improved methods for observing CP violation in B+- ---> K D and measuring the CKM phase gamma,''
 % Phys.\ Rev.\  {\bf D63}, 036005 (2001).
 % [hep-ph/0008090].


%\cite{Giri:2003ty}
%\bibitem{Giri:2003ty}
  %A.~Giri, Y.~Grossman, A.~Soffer, J.~Zupan,
  %``Determining gamma using B+- ---> DK+- with multibody D decays,''
 % Phys.\ Rev.\  {\bf D68}, 054018 (2003).
  %[hep-ph/0303187].


  %\cite{Wang:2012ie}
\bibitem{Wang:2012ie}
  W.~Wang,
  %``CP violation effects on the measurement of $\gamma$ from $B\to DK$,''
  Phys.\ Rev.\ Lett.\  {\bf 110}, 061802 (2013)
  [arXiv:1211.4539 [hep-ph]];
  %%CITATION = ARXIV:1211.4539;%%
  %3 citations counted in INSPIRE as of 01 May 2013
%\cite{Martone:2012nj}
%\bibitem{Martone:2012nj}
  M.~Martone and J.~Zupan,
  %``B^\pm -> D K^\pm with direct CP violation in charm,''
  Phys.\ Rev.\ D {\bf 87}, 034005 (2013)
  [arXiv:1212.0165 [hep-ph]];
  %%CITATION = ARXIV:1212.0165;%%
  %2 citations counted in INSPIRE as of 01 May 2013
  %\cite{Bhattacharya:2013vc}
%\bibitem{Bhattacharya:2013vc}
  B.~Bhattacharya, D.~London, M.~Gronau and J.~L.~Rosner,
  %``Shift in weak phase $\gamma$ due to CP asymmetries in $D$ decays to two pseudoscalar mesons,''
  Phys.\ Rev.\ D {\bf 87}, 074002 (2013)
  [arXiv:1301.5631 [hep-ph]].
  %%CITATION = ARXIV:1301.5631;%%
  %1 citations counted in INSPIRE as of 01 May 2013

\bibitem{CKMfitter}
J. Charles {\it et al.} [
CKMfitter Group], Eur. Phys. J. C{\bf 41}, 1 (2005)
%[hep-ph/0406184]
, updated results and plots available at: http://ckmfitter.in2p3.fr.


%\cite{Browder:2008em}
%\bibitem{Browder:2008em}
%  T.~E.~Browder, T.~Gershon, D.~Pirjol, A.~Soni, J.~Zupan,
  %``New Physics at a Super Flavor Factory,''
 % Rev.\ Mod.\ Phys.\  {\bf 81}, 1887 (2009).
%  [arXiv:0802.3201 [hep-ph]].



%\cite{Wang:2011zw}
\bibitem{Wang:2011zw}
  W.~Wang,
  %``Determining CP violation angle $\gamma$ with B decays into a scalar/tensor meson,''
  Phys.\ Rev.\ D {\bf 85}, 051301 (2012)
  [arXiv:1110.5194 [hep-ph]];%\cite{Wang:2012jba}
%\bibitem{Wang:2012jba}
 % W.~Wang,
  %``B decays into a scalar/tensor meson in pursuit of determining the CKM angle $\gamma$,''
  AIP Conf.\ Proc.\  {\bf 1492}, 117 (2012)
  [arXiv:1209.1244 [hep-ph]];
  %%CITATION = ARXIV:1209.1244;%%
  %1 citations counted in INSPIRE as of 02 May 2013 For earlier discussions, see
  \bibitem{Diehl}
  M.~Diehl and G.~Hiller,
  %``New ways to explore factorization in b decays,''
  JHEP {\bf 0106}, 067 (2001);
  %``Yet another way to measure gamma,''
  Phys.\ Lett.\ B {\bf 517}, 125 (2001).

  %%CITATION = ARXIV:1110.5194;%%


%\cite{Li:1994iu}
\bibitem{Li:1994iu}
  H.~-n.~Li and H.~-L.~Yu,
  %``Perturbative QCD analysis of B meson decays,''
  Phys.\ Rev.\ D {\bf 53}, 2480 (1996)
  [hep-ph/9411308].
  %%CITATION = HEP-PH/9411308;%%

%\cite{Keum:2000wi}
\bibitem{Keum:2000wi}
  Y.~Y.~Keum, H.~-N.~Li and A.~I.~Sanda,
  %``Fat penguins and imaginary penguins in perturbative QCD,''
  Phys.\ Lett.\ B {\bf 504}, 6 (2001)
  [hep-ph/0004004];
  %``Penguin enhancement and B ---> K pi decays in perturbative QCD,''
  Phys.\ Rev.\ D {\bf 63}, 054008 (2001)
  [hep-ph/0004173].
  %%CITATION = HEP-PH/0004173;%%

%\cite{Lu:2000em}
\bibitem{Lu:2000em}
  C.~-D.~Lu, K.~Ukai and M.~-Z.~Yang,
  %``Branching ratio and CP violation of B ---> pi pi decays in perturbative QCD approach,''
  Phys.\ Rev.\ D {\bf 63}, 074009 (2001)
  [hep-ph/0004213].
  %%CITATION = HEP-PH/0004213;%%




%\cite{Li:2012nk}
\bibitem{Li:2012nk}
  H.~-n.~Li, Y.~-L.~Shen and Y.~-M.~Wang,
  %``Next-to-leading-order corrections to $B \to \pi$ form factors in $k_T$ factorization,''
  Phys.\ Rev.\ D {\bf 85}, 074004 (2012)
  [arXiv:1201.5066 [hep-ph]];
  %``Resummation of rapidity logarithms in $B$ meson wave functions,''
  JHEP {\bf 1302}, 008 (2013)
  [arXiv:1210.2978 [hep-ph]].
  %%CITATION = ARXIV:1201.5066;%%
  %7 citations counted in INSPIRE as of 01 May 2013

%\cite{Wang:2012ab}
\bibitem{Wang:2012ab}
  W.~-F.~Wang and Z.~-J.~Xiao,
  %``The semileptonic decays $B/B_s \to (\pi, K)(\ell^+\ell^-,\ell\nu,\nu\bar{\nu})$ in the perturbative QCD approach beyond the leading-order,''
  Phys.\ Rev.\ D {\bf 86}, 114025 (2012)
  [arXiv:1207.0265 [hep-ph]];
    Y.~-Y.~Fan, W.~-F.~Wang and Z.~-J.~Xiao,
  %``Semileptonic decays $B \to D^{(*)} l\nu$ in the perturbative QCD factorization approach,''
  arXiv:1301.6246 [hep-ph].
  %%CITATION = ARXIV:1301.6246;%%
  %%CITATION = ARXIV:1207.0265;%%
  %5 citations counted in INSPIRE as of 01 May 2013

%\cite{Kurimoto:2002sb,Keum:2003js,Lu:2002iv,Li:2003wg,Lu:2003xcLi:2008ts,Zou:2009zza}

%\cite{Kurimoto:2002sb}
\bibitem{Kurimoto:2002sb}
  T.~Kurimoto, H.~-n.~Li and A.~I.~Sanda,
  %``B ---> D(*) form-factors in perturbative QCD,''
  Phys.\ Rev.\ D {\bf 67}, 054028 (2003)
  [hep-ph/0210289].
  %%CITATION = HEP-PH/0210289;%%

\bibitem{Keum:2003js}
  Y.~-Y.~Keum, T.~Kurimoto, H.~N.~Li, C.~-D.~Lu, A.~I.~Sanda,
  %``Nonfactorizable contributions to B ---> D**(*) M decays,''
  Phys.\ Rev.\  {\bf D69}, 094018 (2004)
  [hep-ph/0305335].

%\cite{Lu:2002iv}
\bibitem{Lu:2002iv}
  C.~-D.~Lu, K.~Ukai,
  %``Branching ratios of B ---> D(s) K decays in perturbative QCD approach,''
  Eur.\ Phys.\ J.\  {\bf C28}, 305-312 (2003)
  [hep-ph/0210206].

%\cite{Li:2003wg}
\bibitem{Li:2003wg}
  Y.~Li, C.~-D.~Lu,
  %``Study of pure annihilation type decays B ---> D*(s) K,''
  J.\ Phys.\ G {\bf G29}, 2115-2124 (2003)
  [hep-ph/0304288].

%\cite{Lu:2003xc}
\bibitem{Lu:2003xc}
  C.~-D.~Lu,
  %``Study of color suppressed modes B0 ---> anti-D*0 eta-prime,''
  Phys.\ Rev.\  {\bf D68}, 097502 (2003)
  [hep-ph/0307040].

%\cite{Li:2008ts}
\bibitem{Li:2008ts}
  R.~-H.~Li, C.~-D.~Lu, H.~Zou,
  %``The B(B(s)) ---> D(s) P, D(s) V, D*(s) P and D*(s) V decays in the perturbative QCD approach,''
  Phys.\ Rev.\  {\bf D78}, 014018 (2008)
  [arXiv:0803.1073 [hep-ph]].


%\cite{Zou:2009zza}
\bibitem{Zou:2009zza}
  H.~Zou, R.~-H.~Li, X.~-X.~Wang, C.~-D.~Lu,
  %``The CKM suppressed B(B(s)) ---> anti-D(s)P, anti-D(s)V, anti-D*(s)P, anti-D*(s)V decays in perturbative QCD approach,''
  J.\ Phys.\ G {\bf G37}, 015002 (2010)
  [arXiv:0908.1856 [hep-ph]].

  %\cite{Li:2009xf}
\bibitem{Li:2009xf}
  R.~-H.~Li, X.~-X.~Wang, A.~I.~Sanda and C.~-D.~Lu,
  %``Decays of $B$ meson to two charmed mesons,''
  Phys.\ Rev.\ D {\bf 81}, 034006 (2010)
  [arXiv:0910.1424 [hep-ph]].
  %%CITATION = ARXIV:0910.1424;%%


%\cite{Li:2003yj}
\bibitem{Li:2003yj}
  H.~-n.~Li,
  %``QCD aspects of exclusive B meson decays,''
  Prog.\ Part.\ Nucl.\ Phys.\  {\bf 51}, 85 (2003)
  [hep-ph/0303116].
  %%CITATION = HEP-PH/0303116;%%


  %\cite{Lu:2002ny}
\bibitem{Lu:2002ny}
  C.~-D.~Lu and M.~-Z.~Yang,
  %``B to light meson transition form-factors calculated in perturbative QCD approach,''
  Eur.\ Phys.\ J.\ C {\bf 28}, 515 (2003)
  [hep-ph/0212373].
  %%CITATION = HEP-PH/0212373;%%



%\cite{Cheng:2005nb}
\bibitem{Cheng:2005nb}
  H.~-Y.~Cheng, C.~-K.~Chua, K.~-C.~Yang,
  %``Charmless hadronic B decays involving scalar mesons: Implications to the nature of light scalar mesons,''
  Phys.\ Rev.\  {\bf D73}, 014017 (2006)
  [hep-ph/0508104]; %\cite{Cheng:2013fba}
%\bibitem{Cheng:2013fba}
  H.~-Y.~Cheng, C.~-K.~Chua, K.~-C.~Yang and Z.~-Q.~Zhang,
  %``Revisiting charmless hadronic B decays to scalar mesons,''
  arXiv:1303.4403 [hep-ph].
  %%CITATION = ARXIV:1303.4403;%%

%\cite{Lu:2006fr}
\bibitem{Lu:2006fr}
  C.~-D.~Lu, Y.~-M.~Wang and H.~Zou,
  %``Twist-3 distribution amplitudes of scalar mesons from QCD sum rules,''
  Phys.\ Rev.\ D {\bf 75}, 056001 (2007)
  [hep-ph/0612210].
  %%CITATION = HEP-PH/0612210;%%

%\cite{Han:2013zg}
\bibitem{Han:2013zg}
  H.~-Y.~Han, X.~-G.~Wu, H.~-B.~Fu, Q.~-L.~Zhang and T.~Zhong,
  %``Twist-3 Distribution Amplitudes of Scalar Mesons within the QCD Sum Rules and Its Application to the $B \to S$ Transition Form Factors,''
  arXiv:1301.3978 [hep-ph].
  %%CITATION = ARXIV:1301.3978;%%
  %2 citations counted in INSPIRE as of 01 May 2013


%\cite{Li:2008tk}
\bibitem{Li:2008tk}
  R.~H.~Li, C.~D.~Lu, W.~Wang and X.~X.~Wang,
  %``$B\to S$ Transition Form Factors in the PQCD approach,''
  Phys.\ Rev.\ D {\bf 79}, 014013 (2009)
 [arXiv:0811.2648 [hep-ph]].
  %%CITATION = ARXIV:0811.2648;%%


%\cite{Cheng:2010hn}
\bibitem{Cheng:2010hn}
  H.~-Y.~Cheng, Y.~Koike, K.~-C.~Yang,
  %``Two-parton Light-cone Distribution Amplitudes of Tensor Mesons,''
  Phys.\ Rev.\  {\bf D82}, 054019 (2010)
  [arXiv:1007.3541 [hep-ph]].



%\cite{Gamiz:2009ku}
%\bibitem{Gamiz:2009ku}
%  E.~Gamiz {\it et al.} [ HPQCD Collaboration ],
  %``Neutral $B$ Meson Mixing in Unquenched Lattice QCD,''
 % Phys.\ Rev.\  {\bf D80}, 014503 (2009).
  %[arXiv:0902.1815 [hep-lat]].


%\cite{Neil:2011ku}
\bibitem{Neil:2011ku}
%  E.~T.~Neil    {\it et al.},
  E.~T.~Neil {\it et al.}  [Fermilab Lattice and MILC Collaborations],
  %``B and D meson decay constants from 2+1 flavor improved staggered simulations,''
  PoS LATTICE {\bf 2011}, 320 (2011)
  [arXiv:1112.3978 [hep-lat]].
  %%CITATION = ARXIV:1112.3978;%%
  %15 citations counted in INSPIRE as of 01 May 2013
  %%CITATION = ARXIV:1112.3978;%%



 %\cite{Nakamura:2010zzi}
\bibitem{Nakamura:2010zzi}
%\cite{Beringer:1900zz}
%\bibitem{Beringer:1900zz}
  J.~Beringer {\it et al.}  [Particle Data Group Collaboration],
  %``Review of Particle Physics (RPP),''
  Phys.\ Rev.\ D {\bf 86}, 010001 (2012).
  %%CITATION = PHRVA,D86,010001;%%
  %1330 citations counted in INSPIRE as of 01 May 2013
  %%CITATION = JPHGB,G37,075021;%%


%\cite{Wang:2006ria,Wang:2010ni}
\bibitem{Wang:2006ria}
  W.~Wang, Y.~-L.~Shen, Y.~Li and C.~-D.~Lu,
  %``Study of scalar mesons f0(980) and f0(1500) from B ---> f0(980) K and B ---> f0(1500) K Decays,''
  Phys.\ Rev.\ D {\bf 74}, 114010 (2006)
  [hep-ph/0609082];
  Y.~-L.~Shen, W.~Wang, J.~Zhu and C.~-D.~Lu,
  %``Study of K*0(1430) and a0(980) from B ---> K*0(1430) pi and B ---> a0(980)K Decays,''
  Eur.\ Phys.\ J.\ C {\bf 50}, 877 (2007)
  [hep-ph/0610380];
    X.~Liu, Z.~-Q.~Zhang and Z.~-J.~Xiao,
  %``B ---> K(0)*(1430) eta-prime decays in the pQCD approach,''
  Chin.\ Phys.\ C {\bf 34}, 157 (2010)
  [arXiv:0904.1955 [hep-ph]];
  %%CITATION = ARXIV:0904.1955;%%
  %9 citations counted in INSPIRE as of 02 May 2013
  C.~S.~ Kim, Y.~Li and W.~Wang,
  %``Study of Decay Modes B ---> K(0)*(1430) phi,''
  Phys.\ Rev.\ D {\bf 81}, 074014 (2010)
  [arXiv:0912.1718 [hep-ph]];
    X.~Liu and Z.~-J.~Xiao,
  %``B ---> K*0(1430) K decays in perturbative QCD approach,''
  Commun.\ Theor.\ Phys.\  {\bf 53}, 540 (2010)
  [arXiv:1004.0749 [hep-ph]];
    X.~Liu, Z.~-J.~Xiao and Z.~-T.~Zou,
  %``Charmless hadronic $B_q to K_0^*(1430) \bar{K}_0^*(1430)$ decays in the pQCD approach,''
  arXiv:1105.5761 [hep-ph];
    Z.~-Q.~Zhang,
  %``Study of scalar meson $f_0(980)$ and $K_0^*(1430)$ from $B \to f_0(980)\rho(\omega, \phi)$ and $B \to K^*_0(1430)\rho(\omega)$ Decays,''
  Phys.\ Rev.\ D {\bf 82}, 034036 (2010)
  [arXiv:1006.5772 [hep-ph]].
  %%CITATION = ARXIV:1006.5772;%%
  %11 citations counted in INSPIRE as of 02 May 2013
  %%CITATION = ARXIV:1105.5761;%%
  %3 citations counted in INSPIRE as of 02 May 2013
  %%CITATION = ARXIV:1004.0749;%%
  %4 citations counted in INSPIRE as of 02 May 2013
  %%CITATION = ARXIV:0912.1718;%%
  %%CITATION = HEP-PH/0610380;%%
  %%CITATION = HEP-PH/0609082;%%


%\cite{Wang:2010ni}
\bibitem{Wang:2010ni}
  W.~Wang,
  %``B to tensor meson form factors in the perturbative QCD approach,''
  Phys.\ Rev.\  {\bf D83}, 014008 (2011)
  [arXiv:1008.5326 [hep-ph]].


%\cite{Zou}
\bibitem{Zou}
  Z.~-T.~Zou, X.~Yu and C.~-D.~Lu,
  %``Nonleptonic two-body charmless B decays involving a tensor meson in the Perturbative QCD Approach,''
  Phys.\ Rev.\ D {\bf 86}, 094015 (2012)
  [arXiv:1203.4120 [hep-ph]];
%    Z.~-T.~Zou, X.~Yu and C.~-D.~Lu,
  %``The $B(B_{s})\rightarrow D_{(s)}(\bar{D}_{(s)}) T$ and $D_{(s)}^{*}(\bar{D}_{(s)}^{*})T$ Decays in Perturbative QCD Approach,''
  Phys.\ Rev.\ D {\bf 86}, 094001 (2012)
  [arXiv:1205.2971 [hep-ph]].
  %%CITATION = ARXIV:1205.2971;%%
  %6 citations counted in INSPIRE as of 01 May 2013
  %%CITATION = ARXIV:1203.4120;%%
  %5 citations counted in INSPIRE as of 01 May 2013


%\cite{Freddy:2013apa}
\bibitem{Freddy:2013apa}
  S.~Freddy, C.~S.~Kim, R.~H.~Li and Z.~T.~Zou,
  %``Charmless $B_{u,d,s}\to VT$ decays in perturbative QCD approach,''
  arXiv:1303.6036 [hep-ph].
  %%CITATION = ARXIV:1303.6036;%%


%\cite{Wang:2012ie}
\bibitem{Wang:2012ie}
  W.~Wang,
  %``CP violation effects on the measurement of $\gamma$ from $B\to DK$,''
  Phys.\ Rev.\ Lett.\  {\bf 110}, 061802 (2013)
  [arXiv:1211.4539 [hep-ph]].
  %%CITATION = ARXIV:1211.4539;%%
  %3 citations counted in INSPIRE as of 21 May 2013


%\cite{Martone:2012nj}
\bibitem{Martone:2012nj}
  M.~Martone and J.~Zupan,
  %``B^\pm -> D K^\pm with direct CP violation in charm,''
  Phys.\ Rev.\ D {\bf 87}, 034005 (2013)
  [arXiv:1212.0165 [hep-ph]].
  %%CITATION = ARXIV:1212.0165;%%
  %2 citations counted in INSPIRE as of 21 May 2013

%\cite{Bhattacharya:2013vc}
\bibitem{Bhattacharya:2013vc}
  B.~Bhattacharya, D.~London, M.~Gronau and J.~L.~Rosner,
  %``Shift in weak phase $\gamma$ due to CP asymmetries in $D$ decays to two pseudoscalar mesons,''
  Phys.\ Rev.\ D {\bf 87}, 074002 (2013)
  [arXiv:1301.5631 [hep-ph]].
  %%CITATION = ARXIV:1301.5631;%%
  %1 citations counted in INSPIRE as of 21 May 201

\end{thebibliography}
\end{document}